И. З. ШКУРЧЕНКО

# МЕХАНИКА ЖИДКОСТИ И ГАЗА, ИЛИ МЕХАНИКА БЕЗЫНЕРТНОЙ МАССЫ II.

Во второй части монографии «Механика жидкости и газа, или механика безынертной массы II» рассматривается приложение теории идеальной жидкости к реальным жидкостям и газам. Она содержит восемь примеров практического применения теории безынертной массы, конструктивную схему прибора для измерения динамических сил давления, расчёт усовершенствованной формы крыльчатки центробежного насоса.

В 2007 году редактор поместил примеры практического применения в архиве отдельно. Теперь все эти примеры вошли в текст монографии, что соответствует оригиналу.

Монография адресована инженерам и специалистам в области теоретической и практической гидродинамики и смежных наук.

I. Z. SHKURCHENKO

# MECHANICS OF LIQUID AND GAS, OR MECHANICS OF THE INERTLESS MASS II.

The second part of the monograph "Mechanics of liquid and gas, or mechanics of inertless mass" considers the application of the theory of ideal fluid to real fluids and gases. It contains eight examples of the practical application of the theory of inertial mass, a design diagram of a device for measuring dynamic pressure forces, and a calculation of an improved form of a centrifugal pump impeller. In 2007, the editor archived the case studies separately. Now all these examples are included in the text of the monograph, which corresponds to the original. The monograph is addressed to engineers and specialists in the field of theoretical and practical hydrodynamics and related sciences.

## СОДЕРЖАНИЕ ЧАСТИ I

## ОГЛАВЛЕНИЕ ЧАСТИ II

## Г Л А В А VI. СПОСОБЫ ЗАМЕРА ХАРАКТЕРИСТИК ПОТОКА ЖИДКОСТИ, НАХОДЯЩЕГОСЯ В ДВИЖЕНИИ И В СОСТОЯНИИ ПОКОЯ

В предыдущих разделах мы получили необходимые зависимости и положения, которые полностью характеризуют любое состояние идеальной жидкости, будь то состояние покоя или движения. Окончательной проверкой правильности этих зависимостей служит практика, эксперимент. В этом случае исследование и контроль движения жидкости ведётся с помощью замера необходимых характеристик потока. В характеристики потока входят:

1. геометрические размеры потока,
2. расход массы в единицу времени $M$,
3. линейная скорость $W$,
4. плотность жидкости $\rho$,
5. статические и динамические силы давления $P_{ст}$ и $P_{дин}$,
6. потенциальная и кинетическая энергии $U_п$ и $U_к$,
7. работа $L$,
8. мощность $N$.

Мы назвали полный комплекс характеристик, которыми можно охарактеризовать любое состояние идеальной жидкости, но для каждого её вида движения применима только их часть. Поэтому способы замера этих характеристик рассмотрим применительно к видам движения жидкости.

Сначала познакомимся с приборами и датчиками, которыми контролируются характеристики потока жидкости. Геометрические размеры потока контролируются разными типами приборов и инструментов, предназначенными для этих целей. Перечислять их мы здесь не будем, т. к. они хорошо известны.

Расход массы в единицу времени замеряется специальными расходомерами, либо определяется косвенно, в зависимости от других характеристик потока. Линейная скорость контролируется косвенно, в зависимости от других характеристик потока. Плотность жидкости определяется весовым или каким-либо другим общеизвестным способом. Статические и динамические силы давления контролируются приборами, которые имеют своим рабочим органом плоскость или поверхность. Силы давления являются той основной характеристикой, которая поддаётся непосредственному контролю в процессе движения жидкости. Поэтому остановимся на ней более подробно.

В настоящее время имеется много различных типов подобных датчиков, которые обладают тем или иным техническим несовершенством по отношению к способам контроля сил давления. Это несовершенство выражается в том, что по техническим причинам невозможно создать универсальный датчик сил давления, который был бы применим для всех условий контроля любых сил давления.

По условиям контроля силы давления можно разделить на три группы:

а) статические силы давления постоянные во времени;
б) динамические силы давления постоянные во времени;
в) силы давления переменные во времени.

Это тоже сравнительно общие условия контроля, которые в данном случае определяются только временным фактором. Каждая такая группа контроля сил давления требует определённой схемы датчиков давления, присущей только каждой из этих групп. По временному фактору датчики и приборы для замера сил давления можно разделить на две группы: одни из них замеряют только постоянные во времени силы давления, другие – только переменные во времени силы давления. Теперь рассмотрим приборы и датчики для замера сил давления по отношению к трём группам контроля сил давления и их конструктивным особенностям.

Статические силы давления постоянные во времени замеряются приборами манометрического типа. На этих приборах мы остановимся более подробно при рассмотрении замера потенциальной энергии, так как прямым назначением этих приборов является замер потенциальной энергии жидкости. Функции замера потенциальной энергии и сил давления в манометрических приборах совмещаются следующим образом: рабочим органом этих приборов является объём, который заполняется жидкостью. В связи с тем, что энергия является объёмной единицей измерения, то приборы манометрического типа являются приборами для измерения энергии, но в то же время потенциальная энергия проявляет себя непосредственно через статические силы давления, которые заставляют изменяться форму рабочего органа манометра. По этой причине приборы манометрического типа могут служить одновременно для измерения статических сил давления постоянных во времени и потенциальной энергии.

Правильными, или настоящими, прибором для замера статических сил давления является прибор, рабочим органом которого служит либо поверхность, либо плоскость, т. к. силы давления относятся к плоским единицам измерения. Из этой особенности вытекает конструктивная схема прибора для замера статических сил давления постоянных во времени. Она проста, но подобные приборы не нашли своего применения в практике. Поэтому для замера статических сил давления постоянных во времени здесь приводятся приборы манометрического типа.

Для замера сил давления переменных во времени применяются датчики пьезоэлектрического типа, рабочим органом которых служат кристаллы. На плоскость или поверхность кристаллов действуют замеряемые силы давления. В зависимости от величины этих сил давления в кристалле возникает ток определённой силы, который фиксируется соответствующими регистраторами. Здесь под пьезоэлектрическими датчиками имеются в виду определённые типы приборов, которые работают по схеме пьезоэлектрических датчиков, даже если их рабочим органом не является кристалл. Подобными датчиками, конечно, можно замерять постоянные во времени статические силы давления, но они имеют сравнительно узкий диапазон пределов измерения сил давления.

Для замера динамических сил давления постоянных во времени в современной практике не применяется никаких приборов и датчиков. Поэтому мы дадим принципиальную конструктивную схему подобных приборов и её описание. Изобразим эту схему на рис. 28.

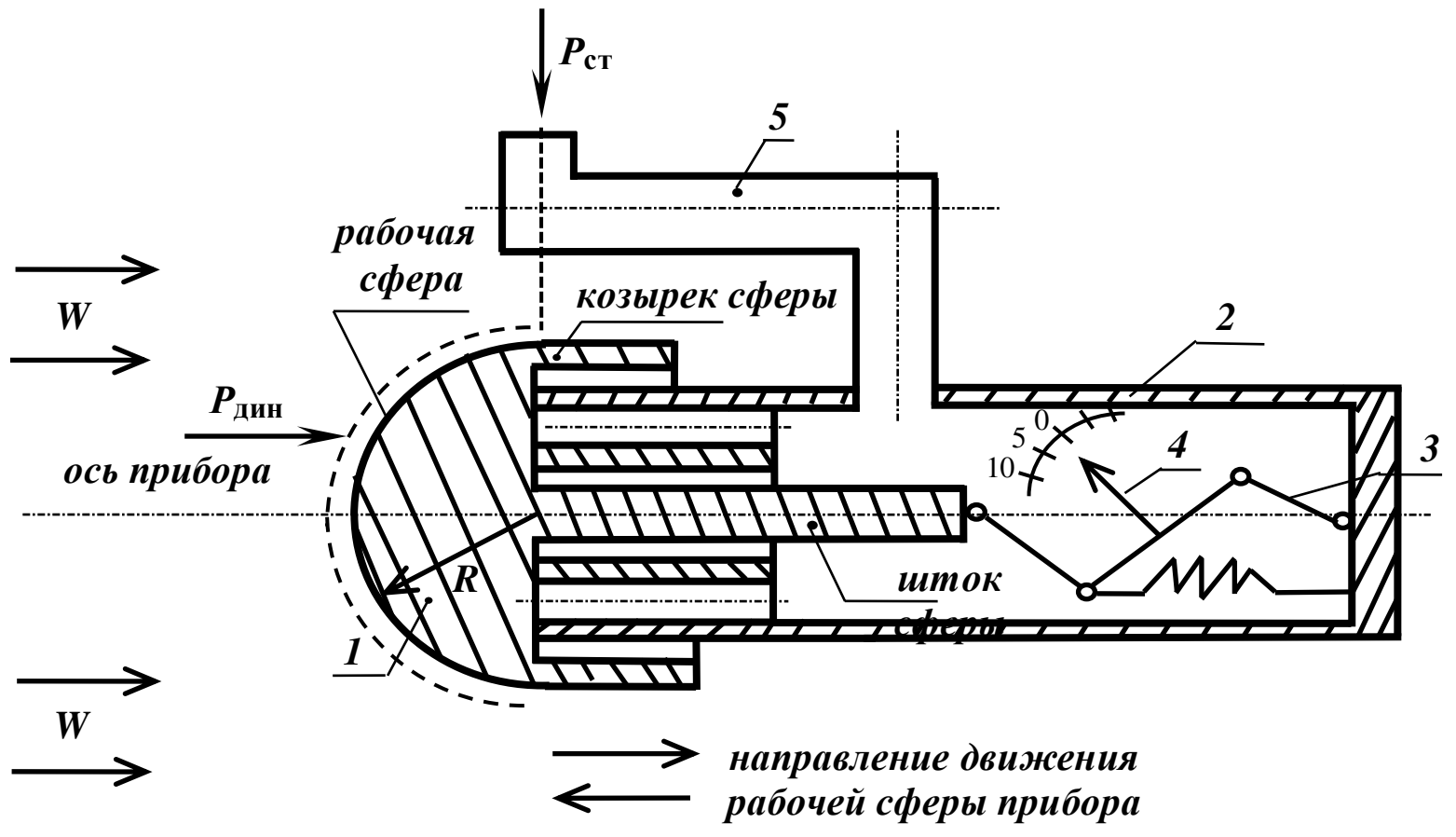

*рис. 28*

Рабочим органом прибора для замера динамических сил давления постоянных во времени является рабочая сфера (*поз.1*), которая имеет непосредственно рабочую площадь $F$, равную половине площади сферы с радиусом $R$ (рис. 28). Конечно, форма рабочей поверхности может быть не только сферой. Почему для данного прибора выбрана сфера, будет показано ниже.

Рабочая сфера (*поз.1*), имеет козырек сферы и шток сферы. С помощью штока она вставляется в гнездо корпуса прибора (*поз.2*), которое позволяет перемещаться рабочей сфере по направлению оси прибора (рис. 28).

Шток соединяется с корпусом прибора с помощью рычажного, пружинного или какого-либо другого механизма (*поз.3*), который даёт возможность определять действующую на шток силу. Этот механизм должен иметь соответствующий регистратор для фиксации сил, действующих на шток, который на рис. 28 отмечен стрелкой со шкалой (*поз. 4*). Внутренняя полость прибора сообщается с потоком с помощью трубки (*поз.5*). Трубка (*поз. 5*) располагается таким образом, чтобы внутренняя полость корпуса прибора заполнялась жидкостью, которая находится под действием статических сил давления $P_{ст}$. Эти статические силы давления должны быть равны статическим силам давления, которые действуют непосредственно на рабочей сфере прибора.

Мы получили принципиальную схему прибора для замера динамических сил давления постоянных во времени. Теперь рассмотрим этот прибор в действии.

Прибор для замера динамических сил давления размещается в таком потоке жидкости, где действуют постоянные во времени динамические силы давления. Он устанавливается таким образом, чтобы ось прибора была параллельна линейной скорости движения потока $W$, а рабочая сфера своей площадью $F$ должна быть направлена навстречу потоку жидкости, как показано на рис. 2. В таком положении прибор закрепляется неподвижно относительно потока жидкости в том месте, где мы желаем произвести замер динамических сил давления потока жидкости.

Со стороны потока на прибор будут действовать статические и динамические силы давления. Статические силы давления одновременно будут действовать на рабочую сферу с двух сторон: со стороны потока на непосредственно рабочую площадь $F$ и со стороны корпуса прибора, т. к. через трубку (*поз. 5*) внутренняя полость его соединена с потоком, который испытывает действие статических сил давления. Тогда действие статических сил давления со стороны потока на площадь рабочей сферы будет уравновешиваться статическими силами давления со стороны корпуса и разность их должна быть равна нулю.

Действие же динамических сил давления со стороны потока на рабочую поверхность $F$ будет уравновешиваться рычажным или пружинным механизмом (*поз. 3*). В соответствии с действием этих сил регистратор (*поз. 4*) зафиксирует величину действующих сил. Затем величину зафиксированных регистратором сил делят на величину рабочей площади сферы, которая равна площади окружности с радиусом $R$. Тогда мы получим величину динамических сил давления, действующих на единицу площади потока. Либо можно сам прибор настроить или сделать таким образом, чтобы он сразу показывал величину динамических сил давления на единицу площади сечения потока. Теперь мы можем замерить динамические силы давления, которые до сих пор никто ещё не замерял.

Мы рассмотрели все способы замера любых сил давления.

Потенциальная энергия замеряется приборами манометрического типа. Выше мы уже рассмотрели способ замера этими приборами потенциальной энергии, т. к. они и предназначены для её замера. Их рабочим органом является объём. Поэтому для этого типа приборов мы принимаем более широкий круг подобных приборов, то есть все приборы, которые имеют своим рабочим органом объём, даже если они имеют другие названия. Сюда входят приборы, принцип действия которых основан на анероидной коробке. В настоящее время подобными приборами замеряют только статические силы давления. Следовательно, проградуировав шкалу этих приборов в соответствии с размерностью потенциальной энергии, ими можно будет делать измерения потенциальной энергии.

Кинетическая энергия, работа и мощность не имеют приборов для непосредственного измерения. Поэтому их контроль ведется косвенным путем, то есть через контроль других характеристик потока.

Теперь рассмотрим измерение характеристик потока жидкости непосредственно по видам движения жидкости.

***Установившийся вид движения жидкости***

С природой установившегося вида движения жидкости мы ознакомились в вышеизложенных разделах. Поэтому нам известно, что характеристики этого потока жидкости определяются в плоскостях исследования, составной частью которых являются площади сечения потока. Отсюда следует, что характеристики потока тоже должны измеряться непосредственно в плоскостях исследования, то есть на площадях сечения потока. Чтобы выполнить это условие, мы должны иметь соответствующие датчики и приборы, габариты которых должны размещаться в плоскости. В то же время мы имеем различные приборы для этих целей, габариты которых выражены в некоторых объёмных единицах. Поэтому подобные приборы занимают в потоке не плоскость, а некоторый объём. По отношению к размерам потока приборов в одних случаях объём может быть практически сопоставим с плоскостью. Тогда приборы показывают результаты одинаковые с расчётными. В других случаях объём того же прибора может существенно менять характеристики потока, т. к. его объём будет изменять в первую очередь геометрические характеристики потока. Тогда эти приборы будут показывать искаженные результаты. Об этих особенностях установившегося потока надо всегда помнить и учитывать их при практических измерениях характеристик потока.

***Плоский установившийся вид движения жидкости***

Особенность этого потока заключается в том, что его характеристики должны замеряться в точках, которые располагаются по линии тока. Поэтому приборы и датчики по своим габаритам практически должны быть соизмеримы с габаритами точки потока. Это значит, что практически добиваются такого совмещения, когда приборы фиксируют результаты без искажения. В этом случае можно считать, что габариты приборов и датчиков не влияют на геометрические характеристики потока. В противном случае необходимо будет делать поправки на искажение. Размещаются приборы и датчики в плоском установившемся потоке в соответствии с радиальными и тангенциальными плоскостями исследования.

***Расходное движение жидкости***

Особенностью этого вида движения жидкости является то, что жидкость втекает или вытекает из объёма, который непрерывно меняет свои границы. В этом случае датчики и приборы должны быть такими, чтобы они не искажали характеристик этого потока. Здесь должны учитываться как габариты потока и датчиков, так и места их установки. Правильность проведенных измерений контролируется по искажениям показаний приборов.

***Акустический вид движения жидкости***

Особенностью данного вида движения является то, что жидкость движется одновременно через поступательную плоскость и нормальную поверхность. При этом границы потока в поступательной плоскости остаются неизменными во время движения жидкости, а высота нормальной поверхности изменяется с течением времени. Это изменение зависит от скорости возмущения или скорости звука. Эти особенности акустического потока должны учитываться в первую очередь при проведении измерений его характеристик. Поэтому различные приборы и датчики должны размещаться непосредственно в поступательной плоскости и на нормальной поверхности исследования. На нормальной поверхности датчики размещаются от поступательной плоскости по направлению движения акустического потока на расстояние не меньше, чем длина волны. При этом габариты датчиков не должны искажать характеристик движения акустического потока. Чтобы сократить число датчиков на нормальной поверхности, нужно вести запись показаний датчиков нормальной и поступательной плоскостей на плоскость фиксации. Эта плоскость должна двигаться со скоростью возмущения или скоростью звука. Либо для её движения должен быть введён коэффициент пропорциональности, полученный относительно скорости возмущения.

Мы получили краткое описание способов замера характеристик потока жидкости. Пожалуй, оно получилось очень кратким и следовало бы его расширить, но для этого нет времени.

## Г Л А В А VII. РЕАЛЬНЫЕ ЖИДКОСТИ И ГАЗЫ

Идеализация реальных жидкостей и газов дала нам возможность получить все необходимые зависимости и положения, которые определяют механические условия существования жидкостей и газов. В отличие от идеальной жидкости реальные жидкости и газы обладают целым рядом свойств, которые не были учтены в механике идеальной жидкости. Реальные жидкости и газы обладают такими свойствами, как вязкость, сжимаемость. Также при изменении состояния реальными жидкостями и газами наблюдается изменение их температуры. Все эти свойства учитываются в науке, которая носит название «термодинамика».

В конечном итоге, нас интересует механика реальных жидкостей и газов. Поэтому мы должны учесть в механике идеальной жидкости свойства реальных жидкостей и газов, чтобы получить их механику. Это положение входит в задачу данного раздела, то есть мы здесь должны указать и согласовать положения механики идеальной жидкости с положениями термодинамики.

### VII. 1. ВЯЗКОСТЬ РЕАЛЬНЫХ ЖИДКОСТЕЙ И ГАЗОВ

Вязкость определяет текучесть реальных жидкостей и газов, вернее, подвижность структурных единиц в жидкостях и газах. В идеальной жидкости для этих структурных единиц мы приняли абсолютную подвижность. В реальных жидкостях и газах эта подвижность далеко не абсолютная. Их вязкость зависит не только от свойств конкретных жидкостей и газов, но и от температуры. Диапазон изменения вязкости реальных жидкостей и газов очень велик, и его надо каким-то образом учитывать. В то же время абсолютная подвижность структурных единиц жидкости и газа есть величина постоянная для каждой конкретной величины плотности $\rho$.

Основной характеристикой движения жидкостей и газов является расход массы в единицу времени. Поэтому изменение в подвижности структурных единиц жидкости и газа в первую очередь отразится на изменении расхода массы в единицу времени.

При абсолютной подвижности структурных единиц расход массы в механике идеальной жидкости имеет для каждого определённого значения плотности $\rho$ однозначное и единственное значение расхода массы в единицу времени. Относительно этого значения определим коэффициент вязкости для любой реальной жидкости или газа. Для этого поступим следующим образом: просто сравним расходы масс реальных жидкостей и газов и идеальной жидкости. Для чего полагаем, что условия для движения идеальной жидкости и реальной одинаковы, то есть геометрические размеры потока, действующие силы давления одинаковы для этих жидкостей. Естественно, что с учётом вязкости мы получим разные результаты для расхода массы в единицу времени. Так как реальные жидкости и газы при любых условиях движения обладают вязкостью, то их расход массы в единицу времени всегда будет меньше расхода массы идеальной жидкости. Вязкость требует дополнительных затрат сил давления. Поэтому расход массы в единицу времени для реальных жидкостей и газов будет меньше расхода массы идеальной жидкости, то есть расход массы идеальной жидкости является тем пределом, которого могут достигнуть реальные жидкости и газы при условии, что подвижность их структурных единиц будет стремиться к абсолютной подвижности. Чтобы определить связь между реальной и идеальной жидкостью, обозначим отношение реального расхода массы в единицу времени к идеальному кинематическим коэффициентом $\mu$, то есть

$$\mu = \frac{M_{\text{р}}}{M_{\text{ид}}}, \tag{1}$$

где $\mu$ – кинематический коэффициент вязкости, $M_{\text{р}}$ – расход массы в единицу времени реальной жидкости, $M_{\text{ид}}$ – расход массы в единицу времени идеальной жидкости.

Кинематическим этот коэффициент назван потому, что расход массы в единицу времени определяется разделом кинематики механики идеальной жидкости.

Мы получили кинематический коэффициент вязкости для реальных жидкостей и газов.

В уравнении (1) расходы масс в единицу времени можно выразить через уравнение движения установившегося вида движения, тогда получим:

$$\mu = \frac{\rho F W_{\text{р}}}{\rho F W_{\text{ид}}} = \frac{W_{\text{р}}}{W_{\text{ид}}}. \tag{2}$$

Мы получили кинематический коэффициент вязкости, выраженный через отношение реальной скорости $W_{\text{р}}$ и идеальной скорости движения $W_{\text{ид}}$.

Для раздела динамики механики жидкости и газа тоже необходим свой динамический, коэффициент вязкости. Раздел динамики определяется силами давления, действующими при движении жидкости и газа. Поэтому на основе сил давления получим динамический коэффициент вязкости.

Жидкость, обладающая вязкостью, требует для своего проталкивания больше сил давления, чем жидкость не имеющая её. Такой жидкостью является идеальная жидкость. Чтобы определить динамический коэффициент вязкости, полагаем, что в потоке при одинаковых геометрических размерах проталкивается расход массы в единицу времени $M_{\text{р}}$, равный расходу массы идеальной жидкости $M_{\text{ид}}$, то есть $M_{\text{р}} = M_{\text{ид}}$. Тогда на проталкивание этого расхода массы потребуется большая величина сил давления $P_{\text{р}}$, чем величина сил давления $P_{\text{ид}}$, которые идут на проталкивание расхода массы идеальной жидкости, то есть $P_{\text{р}} > P_{\text{ид}}$. Используя это условие, запишем динамический коэффициент вязкости через отношение сил давления, получим:

$$\nu = \frac{P_{\text{р}}}{P_{\text{ид}}}. \tag{3}$$

Уравнение (3) выражает динамический коэффициент вязкости.

Мы получили кинематический и динамический коэффициенты вязкости. В отличие от подобных коэффициентов ныне существующей механики жидкости и газа, они не имеют размерности. Практически коэффициенты вязкости получают экспериментальным путём и результаты сводят в таблицы, или для некоторых жидкостей удаётся составить зависимости от температуры.

Применив коэффициенты кинематической и динамической вязкости с зависимостями механики идеальной жидкости, мы сможем получать уже некоторые практические результаты для реальных вязких жидкостей.

## VII. 2. СЖИМАЕМОСТЬ РЕАЛЬНЫХ ЖИДКОСТЕЙ И ГАЗОВ

Сжимаемость является одним из важных свойств реальных жидкостей и газов. Практически все жидкости и газы считаются сжимаемыми. Для механики жидкости и газа различие между жидкостью и газом определяется степенью сжимаемости. Обычно в практических расчётах сжимаемость жидкости не учитывается, а со сжимаемостью газа приходится сталкиваться почти во всех расчётах. В понятие сжимаемости газа входит также его состояние для определённых условий, которые определяются температурой и давлением. Сжимаемость и состояние реальных жидкостей и газов определяются зависимостями и положениями термодинамики, которые в конечном итоге определяют связь между механикой и тепловой энергией.

Это положение даёт нам связь между механикой жидкости и газа и термодинамикой, то есть механика жидкости и газа изучает их механическое состояние и движение, а термодинамика изучает переход тепловой энергии в механическую и наоборот. Таким образом, они дополняют друг друга при практических расчётах различных машин и аппаратов, связанных с использованием механической и тепловой энергии жидкостей и газов.

Эта связь выражается в том, что механика жидкости и газа использует непосредственно зависимости термодинамики. Например, при расчётах движения газа по видам движения, которые определены механикой идеальной жидкости, комплекс зависимостей любого механического вида движения должен быть дополнен термодинамическими зависимостями, которые определяют связь между давлением и плотностью газа, т. к. без этих зависимостей невозможно решить комплекс зависимостей любого механического вида движения газа. Эта связь давления и плотности газа получена в термодинамике при изучении процессов, которые носят следующие названия: изохорный, изобарный, изотермический, адиабатический и политропный.

В зависимости от характера движения газа применяются зависимости того или иного термодинамического процесса. Например, состояние покоя газа можно определить уравнением состояния Клапейрона такого вида:

$$Pv = RT,$$

где $P$ – давление, $v$ – удельный объём, $R$ – газовая постоянная, $T$ – температура.

В то же время термодинамические зависимости по отношению к зависимостям механики идеальной жидкости носят не пассивный характер. Например, в определённых случаях они требуют для видов механического движения даже замены одних механических зависимостей другими. Практические примеры по этому вопросу мы рассмотрим ниже.

Важным положением в термодинамике является положение о работе и энергии, на котором мы должны остановиться более подробно. Рассмотрим работу, которая затрачивается на сжатие определённого объёма газа или жидкости или которая выделяется при расширении тоже определённого объёма газа или жидкости. В термодинамике [12] она записывается таким уравнением:

$$L = \int_{V_1}^{V_2} P dV. \qquad (6)$$

С помощью уравнения (6) можно записать работу, которая либо была затрачена на сжатие, либо которая выделяется при расширении объёма газа. При этом его необходимо будет дополнить зависимостью термодинамического процесса, который соответствует данным условиям движения газа или жидкости. Тогда мы получим соответствующую работу.

Этим же уравнением (6) записывается также энергия сжатия, которая может быть использована как работа при расширении объёма газа из первоначального положения до некоторого второго положения. Следовательно, уравнение (6) в зависимости от назначения может быть либо уравнением энергий, либо уравнением работ.

Это уравнение можно записать по отношению к единице объёма газа или жидкости. Тогда за единицу объёма принимают единицу объёма первоначального положения или состояния газа. В этом случае пределы определённого интеграла уравнения (6) будут выглядеть так:

$$L=\int_{1}^{V_2} P\,dV\,. \qquad (7)$$

Уравнение (7) является уравнением работ или энергии сжатия единицы объёма газа или жидкости.

Теперь определим эти уравнения по отношению к уравнениям работ и энергий механики идеальной жидкости.

Если, например, взять книгу [7], то в ней при рассмотрении механической энергии, то есть при рассмотрении уравнений Бернулли для движения газа, интеграл кинетической энергии такого вида $\int_{p1}^{p2}\frac{dP}{?}$ [1] принимается за уравнение энергии сжатия, вернее, за уравнение термодинамической энергии, если говорить более обобщенно, т. к. в этот интеграл при демонстрации различных случаев движения газов в качестве примера подставляются зависимости либо изотермического процесса, либо адиабатического. Но нигде мы не увидим, чтобы для этих целей использовались уравнения энергий (6) или (7). Это значит, что фактически в уравнении Бернулли кинетическая энергия потока не учитывалась, так как её заменяли энергией сжатия.

Теперь рассмотрим, можно ли применять в уравнении Бернулли зависимости изотермического и адиабатического процессов. Для чего запишем уравнение энергии установившегося вида движения для единицы объёма:

$$U=U_{\text{п}}+\int_{0}^{P_2} dP=\text{const.}$$

Как мы знаем, это есть уравнение Бернулли для потока жидкости. Оно выражает постоянство полной энергии потока $U$, которая состоит из потенциальной энергии $U_{\text{п}}$ и кинетической энергии $\int_{0}^{P_2} dP$. Это уравнение предусматривает постоянство объёма для соответствующего давления, то есть сущность его выражается в произведении сил давления на объём. Поэтому, изменится ли давление или объём односторонне, величина произведения тоже изменится, то есть не будет постоянной величиной. Зависимости адиабатического и изотермического процессов предусматривают изменение объёма установившегося потока. Что, в свою очередь, приведёт к изменению полной механической энергии потока, если применить их непосредственно в уравнении установившегося вида движения. Если же нам по условию термодинамического состояния установившегося потока газа будет необходимо учесть изменение объёма, то мы будем обязаны дополнить уравнение энергии установившегося потока уравнением энергии сжатия (7), чтобы учесть изменение объёма. Тогда оно примет вид:

$$U_{\text{д}}=U_{\text{п}}+\int_{0}^{P_2} dP\pm\int_{1}^{V_2} P\,dV. \qquad (8)$$

---

[1] в оригинале знаменатель неразборчив для редактора

Уравнение энергии сжатия (7) непосредственно учитывает термодинамические условия в уравнении (8). Поэтому потери или прибыль термодинамической энергии мы будем обязаны учитывать уравнением (7) для уравнения (8), т. к. убыль или прибыль тепловой энергии выражается непосредственно зависимостями термодинамических процессов. Поэтому мы должны подставить их зависимости в уравнение ( $\int_1^{V_2} PdV$ ), а не в уравнение кинетической энергии ( $\int_0^{P_2} dP$ ), как это делалось в современной механике жидкости и газа.

Это существенное различие по вопросу энергии между положениями существующей механики жидкости и газа и данной работой, которое очень просто проверяется экспериментом. Сущность расхождения заключается в том, что современная механика жидкости и газа не делает различия по вопросу энергии между положениями механики и термодинамики. Поэтому член уравнения Бернулли, выражающий кинетическую энергию потока, просто выбрасывают и заменяют термодинамической энергией, выраженной через зависимости термодинамических процессов.

Сущность новых положений по этому вопросу заключается в том, что уравнение Бернулли выражает количественно полную механическую энергию установившегося потока жидкости или газа, которая должна быть постоянной для этого потока. А термодинамический член, выражающий работу сжатия (7), добавляется к нему в том случае, когда необходимо учесть либо потери механической энергии, связанные с утечкой тепловой энергии, либо приток механической энергии, связанный с подводом тепла к этому потоку. Тогда утечки или приток тепловой энергии, выраженные через зависимости соответствующих термодинамических процессов, вводят в уравнение Бернулли с помощью термодинамического члена работы сжатия, записанного уравнением (7).

Мы рассмотрели только один случай, связанный с распределением энергии в установившемся виде движения реальных жидкостей и газов. Если мы опять обратимся к положениям современной механики жидкости и газа, то мы не найдем в них раздела, который бы определял, что такое работа, как её определять и какие для неё должны быть зависимости. Все эти понятия полностью взяты из термодинамики. Поэтому для жидкостей работу определяют в основном практическим способом, а для газов берут её как адиабатическую работу.

Рассмотрим понятие работы относительно новых положений механики жидкости и газа. Как мы уже знаем, понятие работы в механике связано с расходным видом движения. Например, энергия, которая должна быть преобразована в работу, транспортируется к месту использования в соответствии с положениями и зависимостями установившегося вида движения жидкостей и газов. Поэтому опять обратимся к уравнению Бернулли. Для реальных жидкостей в соответствии с этим уравнением к месту преобразования подойдет поток жидкости с энергией равной

$$U = U_{\text{п}} + \int_0^{P_2} dP,$$

при условии, что транспортируемые жидкости не отличаются от идеальной, и что не было никаких потерь, связанных с утечкой тепла.

Отметим, что мы ещё не рассматривали вопрос о трении в жидкостях и газах, поэтому здесь не упоминаются потери механической энергии, связанные с этим явлением. Они будут рассмотрены ниже.

В связи с тем, что к месту использования транспортируется реальная жидкость в реальных условиях, то мы должны учесть потери механической энергии, связанные с вязкостью жидкости и термодинамическими условиями движения. Это значит, что мы должны из полной энергии потока $U$ вычесть потери энергии, связанные с вязкостью $\Delta U_{\text{в}}$ и потери энергии, связанные с термодинамическими условиями движения $\Delta U_{\text{т}}$. Тогда получим фактическое, или действительное, количество энергии $U_{\text{д}}$. Оно будет равно:

$$U_{\text{д}} = U - \Delta U_{\text{в}} - \Delta U_{\text{т}}. \qquad (9)$$

Потери энергии, связанные с вязкостью, определяются следующим образом. Эти потери непосредственно относятся к кинетической энергии потока. Для данного вида движения уменьшение происходит за счёт уменьшения расхода массы в единицу времени. Поэтому мы должны умножить кинетическую энергию в уравнении Бернулли на коэффициент кинематической вязкости $\mu$, получим:

$$U_{\text{к.д}} = \mu \int_{0}^{P_2} dP . \tag{10}$$

Равенство (10) выражает действительную кинетическую энергию для реальных вязких жидкостей.

Мы умножили кинетическую энергию на кинематический коэффициент вязкости потому, что расход массы в единицу времени входит в единицу объёма как линейная скорость, которая является длиной единицы объёма. Умножив на $\mu$, мы тем самым изменили длину единицы объёма и сам объём.

Потери энергии, связанные с термодинамическими условиями движения, определяются уравнением энергии сжатия (7), то есть

$$\Delta U_{\text{т.д}} = \int_{1}^{V_2} PdV . \tag{11}$$

Теперь подставим уравнения (10) и (11) в уравнение (9), получим:

$$U_{\text{д}} = U_{\text{п}} + \mu \int_{0}^{P_2} dP - \int_{1}^{V_2} PdV . \tag{12}$$

Уравнение (12) выражает действительную полную энергию потока $U_{\text{д}}$. Оно отличается от уравнения Бернулли тем, что в нём учтены потери, связанные с вязкостью реальных жидкостей, и термодинамические условия движения. Следовательно, уравнение Бернулли является основой для уравнения действительной энергии потока (12).

Уравнение (12) без каких-либо переделок пригодно для установившегося потока жидкости или газа.

Теперь считаем, что поток жидкости или газа подошёл к месту преобразования энергии в работу. Полагаем, что преобразователь энергии в работу имеет своим коэффициентом полезного действия единицу. После преобразования действительной энергии установившегося потока *жидкости* в работу мы получим, что работа $l$ единицы объёма, будет равна действительной энергии установившегося потока жидкости, то есть

$$l = U_{\text{д}} . \tag{13}$$

После преобразования действительной энергии установившегося потока *газа* в работу мы получим, что работа единицы объёма $l$, будет больше действительной энергии установившегося потока газа на величину адиабатического расширения газа от давления потока до давления внешней среды $P_{\text{вн}}$, то есть

$$l = U_{\text{д}} + l_{\text{а}} . \tag{14}$$

В равенстве (14) действительная энергия потока определяется уравнением (12), а работа адиабатического расширения газа определяется уравнением работ сжатия (7) с помощью зависимостей адиабатического процесса. При этом в уравнении (7) предел интеграла $V_2$ надо будет заменить на предел расширения объёма $V_{\text{ср}}$, который соответствует среде, куда происходит расширение газа. Тогда уравнение (7) примет вид:

$$l_{\text{а}} = \int_{1}^{V_{\text{ср}}} PdV. \qquad (15)$$

В уравнении (15) второй предел интеграла как единицу объёма мы оставили без изменения. Это, конечно, не совсем правильно, т. к. вместо единицы сюда следует подставить объём $V_2$, вернее, предел интеграла уравнения (11). Тогда бы мы учли изменение объёма с учётом термодинамических потерь.

Далее следует подставить в уравнение (15) давление, выраженное через зависимости адиабатического процесса, и тогда получим окончательное уравнение адиабатической работы для единицы объёма газа.

Теперь вернёмся к уравнению (14), т. к. оно может показаться прямым нарушением закона сохранения энергии. Такое мнение, конечно, не будет правильным, т.к. здесь нет никакого нарушения этого закона. Просто подобным уравнением учтены особенности движения газов. Мы здесь рассматривали только преобразование энергии установившегося потока газа в работу. Но если бы нам сначала пришлось затратить работу, чтобы создать подобный установившийся поток газа, то мы бы увидели, что эта работа превышает по величине полученную, так как нам пришлось бы покрывать ещё потери энергии. По этой причине, например, когда хотят использовать энергию жидкостей и газов, стремятся создать установившийся поток из жидкостей и затрачивать на него работу, а работу движения получать от газового потока. Так что уравнение (14) просто учитывает это стремление.

Мы рассмотрели второй пример преобразования энергии в работу. Из-за недостатка времени нам, пожалуй, не придётся рассматривать полный комплекс этих случаев. Будем надеяться, что эти два примера дали общую нить для дальнейшего исследования работы и энергии для различных условий движения жидкостей и газов.

Отметим, что в термодинамике, как и в современной механике жидкости и газа, нет чёткого разграничения между давлением и энергией. Поэтому при исследовании термодинамических зависимостей надо быть предельно внимательным.

## Г Л А В А VIII. РАЗНОЕ

### VIII. 1. НЕКОТОРЫЕ ПРИМЕРЫ ПРИМЕНЕНИЯ ЗАВИСИМОСТЕЙ МЕХАНИКИ ИДЕАЛЬНОЙ ЖИДКОСТИ И МЕХАНИКИ РЕАЛЬНЫХ ЖИДКОСТЕЙ И ГАЗОВ

До сих пор мы рассматривали состояние покоя и движения идеальной жидкости без каких-либо ограничений практического характера. Постановка вопроса в таком плане дала нам возможность выявить полный комплекс зависимостей по видам движения и состоянию покоя, которые составляют полное содержание механики жидкости и газа.

Все эти положения и зависимости выражают чисто механические свойства жидкостей и газов. Поэтому идеальная жидкость рассматривалась при постоянной плотности и абсолютной подвижности своих структурных единиц, которые называются молекулами и атомами.

В разделе о реальных жидкостях и газах были учтены те свойства, которые присуще реальным жидкостям и газам, т.е. вязкость, сжимаемость и их состояние в зависимости от термодинамических условий. Все эти свойства не изменили сущности положений и зависимостей механики идеальной жидкости, а лишь расширили её границы до границ реальных жидкостей и газов. Поэтому мы получили механику непосредственно реальных жидкостей и газов.

Практическое использование жидкостей и газов предусматривает ряд условий, которые не были учтены в положениях и зависимостях, как механики идеальной жидкости, так и механики реальных жидкостей и газов.

Условия, связанные с практическим использованием жидкостей и газов, разнообразны и многочисленны. Они составляют содержание целого ряда прикладных наук. Поэтому рассмотреть все условия просто невозможно. Они, например, определяются движением жидкостей и газов в трубопроводах, их состоянием в резервуарах и ёмкостях, изучением морских и воздушных течений, использованием в газовых и гидравлических машинах, движением твёрдых тел различной формы в жидкостях и газах и так далее.

По отношению к этим условиям положения механики реальных жидкостей и газов носят сравнительно частный характер, т.к. они не включают их в себя. Поэтому в данном разделе ограничимся несколькими характерными примерами практического использования зависимостей механики жидкости и газа, чтобы только показать несколько различных вариантов их практического использования и определить связь положений и зависимостей механики жидкости и газа непосредственно с практикой.

ПРИМЕР №1 <…>

*<u>Примечание редактора</u>*: по некоторым причинам редактор перенёс описание примеров практического использования зависимостей механики безынертной массы в *Приложение* к данной редакции текста механики безынертной массы. См.: *Приложение.*

## VIII. 2. СОПРОТИВЛЕНИЕ ЖИДКОСТЕЙ И ГАЗОВ

Само слово «сопротивление» означает - противиться движению. При движении жидкостей и газов существует целый ряд объективных причин, которые создают определённые условия, связанные с помехами движению. В зависимости от этих условий помехи могут быть большими и малыми. По отношению к движущимся жидкостям и газам они выражаются количеством непроизводительно затраченной работы или энергии при их движении. Поэтому количественная сторона сопротивления определяется количеством непроизводительно затраченной энергии.

В целом, сопротивление, как одно из основных понятий механики жидкости и газа, определяется количеством непроизводительно затраченной энергии и причинами этих затрат. По этой причине сопротивление относится к комплексному понятию. Поэтому в тех случаях, когда считают, что причина сопротивления общеизвестна, сопротивления характеризуются просто количеством энергии. Тогда их размерностью является размерность работы. В других случаях сопротивление характеризуют ещё по отношению к причине, например, по отношению к длине потока, тогда размерностью сопротивления станет размерность работы и длины. Иногда сопротивление выражают безразмерной величиной. Тогда эту величину записывают через отношение работ или энергий. К такому методу обычно прибегают тогда, когда в сопротивлении необходимо выразить несколько причин, которые либо неизвестны, либо выражаются в сумме. Примером такого выражения может служить коэффициент полезного действия.[2] Следовательно, для сопротивления непроизводительная затрата энергии является лишь количественной стороной, а причина является для него определяющей стороной.

Рассмотрим движение жидкостей и газов по отношению к причинам сопротивления.

В разделах механики жидкости и газа, касающихся движения идеальной жидкости, мы не встречали таких причин. Это связано с тем, что мы рассматривали идеальные условия движения. При переходе к реальным жидкостям и газам мы сразу столкнулись с ними. Все причины можно разделить на две основные группы. К первой группе отнесём причины, которые связаны непосредственно со свойствами массы реальных жидкостей и газов. К ним относится вязкость и термодинамические условия движения. Вязкость определяется коэффициентами кинематической и динамической вязкости, а термодинамические условия – через уравнения работ и энергии сжатия жидкостей и газов (VII-7). Хотя термодинамические условия и вязкость выражаются различными зависимостями для жидкостей и газов, но, в конечном итоге, с их помощью мы определяем величину нерациональных затрат энергии. Как это делается, мы показали на примерах приведённых выше. Конечно, они далеко не полно отражают все способы их применения в зависимости от условий движения. Изучение этих способов относится непосредственно к разряду прикладных наук. Поэтому мы оставим общие положения для первой группы причин сопротивления.

Ко второй группе относятся причины, которые определяются условиями, возникающими на поверхности контакта твёрдых тел с жидкостями и газами при их движении. Поверхность контакта твёрдых тел, прежде всего, является границей различных потоков. На этих границах они движутся в соответствии с положениями механики жидкости и газа о четырёх видах движения. Поверхность твёрдого тела обладает определённой формой, которая формирует соответствующий поток. В свою очередь, эта поверхность может иметь неровности, связанные, например, с чистотой обработки. Сама поверхность и неровности поверхности в определённых случаях являются причинами потерь энергии потока.

---

[2] в работе «Движение твёрдых тел в жидкостях и газах с точки зрения механики безынертной массы» автор даёт дополнительную новую характеристику – КЭИС, т.е. коэффициент эффективного использования сил.

Форма поверхности как граница потока может влиять на потери его энергии двумя способами. В первом случае потери происходят, когда форма поверхности твёрдого тела создаёт асимметрию потока, например, в его установившемся виде движения. Во втором случае форма поверхности нарушает распределение энергии потока между кинетической и потенциальной. Неровности поверхности влияют на потери энергии только как нарушители распределения энергии потока.

Отметим, что, возможно, на поверхности контакта жидкости и твёрдого тела существует ещё другое силовое взаимодействие между ними, которое не связано с формой поверхности и которое в настоящее время мало изучено. Об этом взаимодействии свидетельствуют капиллярные явления и статическое электричество, возникающее в стенках твёрдых тел при движении жидкостей и газов.

Мы получили полный список условий, при которых возникают потери полной энергии потока. Другие виды потерь в настоящее время нам неизвестны. Таким образом, мы определили две группы причин, которые указывают только на существование сопротивления в жидкостях и газах.

Отметим, что в положениях современной механики жидкости и газа указывается на наличие трения в жидкостях и газах. В связи с тем, что их структурные единицы безынертны, то о существовании трения в жидкостях и газах не может быть и речи. [3]

## VIII. 2. О ЗАКОНЕ СОХРАНЕНИЯ ЭНЕРГИИ

Закон есть закон. Поэтому изменять его никто не в праве. Объективность закона сохранения энергии доказана практикой и не вызывает сомнения. Следовательно, все трудности, связанные с этим законом, заключаются лишь в умении применять его в своей практике. В этом плане рассмотрим закон сохранения энергии.

Основное назначение этого закона сводится к утверждению количественной стороны энергии и к признанию многообразия её форм проявления в природе.

Существование жидкостей и газов в окружающем нас мире определено законом сохранения состояния, согласно которому они сохраняют энергию силового поля. При этом жидкости и газы находятся либо в состоянии покоя, либо в состоянии установившегося движения.

Силовые поля могут быть векторными и скалярными. Действие сил скалярного силового поля определяется непосредственно свойствами массы жидкости и газа. Оно проявляется в их замкнутом объёме при условии, что стенки этого объёма могут воспринимать действие сил давления жидкостей и газов.

Действие сил векторного силового поля не зависит от свойств массы жидкостей и газов. Это силовое поле существует как бы независимо от них. Поэтому жидкости и газы испытывают силовое воздействие векторного поля сил лишь в том случае, когда находятся в сфере его действия. Объём сферы действия этого поля определяется некоторыми природными условиями, которые не зависят от жидкостей и газов. В главе VI мы получили количественные зависимости, которые описывают энергию жидкостей и газов в векторном и скалярном полях сил. Для реальных жидкостей и газов количество энергии зависит ещё от термодинамических условий их существования. Эта часть энергии записывается уравнением (VII-7).

В окружающем нас мире жидкости и газы находятся одновременно под действием скалярных и векторных силовых полей и при определённых термодинамических условиях существования. Поэтому полная энергия этих жидкостей и газов состоит из трех частей, которые мы можем определить раздельно с помощью соответствующих зависимостей, а затем суммировать. Следовательно, жидкости и газы в окружающем нас мире обладают энергией, и закон сохранения к ним применим.

В то же время состояние покоя и движения жидкостей и газов выражается через их структурные единицы, которые называются молекулами и атомами. Когда каждая из всех частиц перемещается, то в сумме это определяет движение жидкости или газа. Когда каждая из всех частиц находится в состоянии покоя, то в сумме они определяют состояние покоя жидкости или газа.

---

[3] структурные единицы среды не могут ни ускоряться, ни замедляться, впоследствии взаимодействие, которое назвали силами трения, должно получить соответствующее название.

Согласно уравнению сил расходного вида движения, каждая структурная единица перемещается от действия любой силы, как бы мала эта сила ни была. При этом скорость движения структурной единицы соответствует величине действующей на неё силы. Поэтому изменение состояния структурных единиц означает изменение действующих на них сил. Например, для изменения состояния структурной единицы жидкости и газа достаточно действия одной силы, а для фиксации её в состоянии покоя необходимо, как минимум, действие двух сил, одна из которых должна действовать в направлении движения, а вторая – против этого движения. Следовательно, любое состояние структурных единиц определяется только действующими на них силами, а не их массой. Поэтому они безынертны.

В связи с тем, что структурные единицы жидкостей и газов, как составные части их массы, не имеют механической энергии и не могут её накапливать, то из этого следует, что закон сохранения энергии к ним не применим. Поэтому они не могут ни подчиняться ему, ни нарушать его. Отсюда также следует, что по отношению к структурным единицам массы и к самой массе жидкостей и газов в целом не приемлемы такие крайние определения, как подчиняются они или не подчиняются закону сохранения энергии. Они просто не входят в сферу его действия.

Теперь мы можем сделать определённые выводы по отношению применения закона сохранения энергии в механике жидкости и газа. В термодинамике, например, жидкости и газы называются рабочим телом. Пожалуй, это наиболее правильный термин, применённый к ним. Действительно, в состоянии жидкости и газа вещество является сравнительно идеальным рабочим телом для преобразования одного вида энергии в другой. Идеальность такого состояния определяется именно тем, что масса жидкостей и газов не обладает собственной механической энергией. Например, в турбинах гидроэлектростанций вода преобразует энергию гравитационного поля Земли в механическую работу. В тепловых электростанциях вода преобразует тепловую энергию, выделяющуюся при сгорании топлива, тоже в механическую работу. Как в этих случаях, так и во многих других при практическом использовании жидкостей и газов не было замечено, чтобы какая-либо часть энергии шла непосредственно на изменение их собственной механической энергии. Поэтому с точки зрения закона сохранения энергии мы должны относиться к жидкостям и газам как к рабочему телу, которое не имеет собственной энергии, и применять этот закон только к непосредственным источникам энергии. Например, к гравитационному полю Земли, к топливу, к движущемуся твёрдому предмету и т.д. Тогда по отношению к этим источникам мы сможем выяснить количественную сторону энергии, которая должна сохраняться или изменяться в соответствии с законом сохранения энергии.

Непосредственное проявление любого количества энергии происходит с помощью массы жидкостей и газов (рабочего тела). Поэтому количество энергии можно выразить также через массу жидкостей и газов. Необходимые для этого зависимости были получены в механике жидкости и газа. В этом выражается её частный характер в сравнении с законом сохранения энергии. Например, поле земного тяготения представляет собой источник определённой формы энергии, количество, или запас, которой мы в настоящее время даже не можем определить. Поэтому, применяя закон сохранения энергии к полю земного тяготения как к источнику энергии, мы можем сказать только, что запасы его энергии очень велики и что они имеют некоторую определённую величину, которую мы не знаем, т. к. природа этого источника нам неизвестна. В то же время человек использует его энергию в своей практической деятельности: строит гидроэлектростанции, приливные электростанции, пользуется морскими и воздушными течениями и т. д. По отношению к общему источнику гравитационной энергии все эти реки, течения, приливы являются частными случаями использования его энергии, которая проявляет себя в виде движения водных масс.

Теперь проследим эту связь. В данном примере вода воспринимает силовое воздействие гравитационного поля Земли и под действием этих же самых сил она совершает движение, в том числе она вращает лопасти турбин гидроэлектростанций. В то же время само гравитационное поле не может вращать турбины. Если лишить воду силового воздействия, она потеряет свою способность вращать турбину. Отсутствие действия гравитационного поля в настоящее время называется невесомостью. Следовательно, вода здесь является промежуточным звеном между источником энергии (гравитационное поле) и преобразователем энергии в работу (турбина), то есть вода является здесь рабочим телом, которое способно воспринимать силовое действие гравитационного поля и передавать его на лопасти турбин. Эта способность воды определяется только безынертностью её массы. В противном случае она бы стала просто потребителем энергии этого источника, как любое твёрдое инертное тело. В связи с тем, что под действием сил гравитационного поля Земли находится объём всего источника рабочего тела (река), то всё это воздействие в сумме определяется как энергия, так как энергия проявляется непосредственно через действие сил. Энергия этого источника записывается зависимостями механики жидкости и газа. Таким образом, происходит превращение источника рабочего тела (реки) в источник энергии, хотя настоящим источником энергии остаётся поле земного тяготения. Но теперь уже к реке, как к источнику энергии, мы в праве применить закон сохранения энергии. В данном случае нас будет интересовать его количественная сторона. Определяя её, мы можем определить производительные и непроизводительные затраты энергии, например, коэффициент полезного действия турбин.

Мы выяснили, что река является одновременно источником рабочего тела и источником энергии. Теперь определим значение турбины по отношению к этому источнику. Турбина относится к разряду лопастных машин, двигателей. В данном случае работа этой машины основана на использовании энергии реки. Сама река как источник рабочего тела может быть вечной или невечной, т. к. любая река может иссякнуть, переменить русло, наконец, её можно перегородить плотиной. Тогда по отношению к конкретно расположенной турбине река в качестве источника рабочего тела не является вечной. В то же время река в качестве источника энергии получает её от гравитационного поля Земли. По отношению к существованию человечества источник гравитационной энергии является вечным источником, т. к. при его исчезновении человечество прекратит своё существование. Следовательно, источник гравитационной энергии можно рассматривать как вечный. Поэтому турбину, работа которой основана на использовании его энергии, можно отнести к разряду вечных. Это значит, что подобные турбины относятся к разряду вечных двигателей третьего рода. В данном случае вечен, конечно, не сам двигатель, а вечен источник энергии.

Вот, пожалуй, и всё, что можно сказать о законе сохранения энергии. Конечно, следовало бы написать несколько больше, т. к. закон сохранения энергии и механика жидкости и газа этого заслуживают. В настоящее время, пожалуй, это не принципиально, т. к. приходится ждать, когда настоящая механика жидкости и газа понадобится кому-либо.

<div align="right">Отпечатано: 9 ноября 1971 года.<br>Перепечатано с сокращениями: 15 ноября 1977 года.[4]</div>

[4] Подготовка рукописи к печати началась с сокращённого варианта, сделанного автором в 1977 году. Данный текст – полный.

## ЛИТЕРАТУРА

[5] Примечания ко всем заявкам даны автором

## ПРИЛОЖЕНИЕ

*Как читатель заметил, механика безынертной массы требует от него переосмысления положений существующей гидромеханики, начиная с уравнения Бернулли. Поскольку критерием истины является практика, то помочь переосмыслению может только руководство положениями механики безынертной массы в практической деятельности, для чего автор приводит восемь примеров применения зависимостей механики безынертной массы.*

*Впоследствии автором были созданы три работы, в которых тема прикладного применения зависимостей механики безынертной массы получила развитие. Таким образом, автор положил начало развитию прикладной механики безынертной массы. Поскольку автор в дальнейших прикладных исследованиях изменил некоторые выводы, которые в «Примерах» даны в первом приближении, то редактор считает, что для полного понимания новой теории необходимо ознакомится с остальными работами автора, прежде чем попробовать применить зависимости на практике.*

*Как более проработанный пример применения положений и зависимостей механики безынертной массы, наряду с «Примерами», редактор помещает в Приложении принцип конструирования крыльчатки центробежного насоса, который был дан автором во второй работе - «Строение Солнца и планет солнечной системы с точки зрения механики безынертной массы» (1973 -1974 г.г.). В этой работе автор дополнил положения о плоском установившемся виде движения.*

*В следующей работе «Движение твёрдых тел в жидкостях и газах с точки зрения механики безынертной массы», в частности, дан принцип конструирования гребного винта (движителя) и принцип механизма машущего полёта.*

*Теория автора легко проверяется на практике. Полвека никто не пробовал проверить.*

*Приведу отрывок из работы «Движения твёрдых тел в жидкостях и газах», в котором автор подводит определённый итог проделанной им исследовательской работы:*

Мы закончили ещё одну работу по прикладному применению законов механики безынертной массы. Первой работой, написанной в этом плане, была работа под названием «Строение Солнца и планет солнечной системы с точки зрения механики безынертной массы», где было установлено внутреннее строение Солнца и планет солнечной системы, дан расчёт центробежных насосов и компрессоров.

В данной работе мы определили понятие обтекаемости твёрдых тел при их движении в жидкостной и газовой среде, установили геометрические размеры потоков взаимодействия на лобовой и тыльной поверхностях твёрдого тела, выяснили очень важный вопрос, что газы в потоках взаимодействия, при любых скоростях движения твёрдого тела, ведут себя как несжимаемые жидкости, в соответствии с изохорным процессом в термодинамике. Далее мы выяснили, что жидкости тоже подчиняются изохорному процессу, но ведут себя с определёнными особенностями, присущими только жидкостям. В термодинамике процессы для жидкостей не даны. В общем, до настоящего времени никто не знал, куда можно было бы приспособить изохорный процесс. Мы этот вопрос выяснили и установили, что он присутствует и при акустическом виде движения, и в потоках взаимодействия, и в сопле Лаваля, и в движителях, и в различных лопастных машинах. В связи с этим мы должны сделать уточнение. В работе «Строение Солнца и планет солнечной системы с точки зрения механики безынертной массы», следуя общепринятому мнению, мы написали, что газы в колесах центробежных компрессоров ведут себя в соответствии с адиабатическим процессом. После проведенных в данной работе исследований, это положение надо считать неверным, т. к. мы установили, что в потоках взаимодействия газы подчиняются изохорному процессу. По этой причине газы в колесах центробежных компрессоров тоже будут подчиняться изохорному процессу. Эта поправка очень существенна, и с ней нельзя не считаться.

Также мы выявили необходимость такой новой технической характеристики, как коэффициент эффективного использования сил (КЭИС). В настоящее время для подобных целей существует только коэффициент полезного действия (КПД). Он характеризует эффективность использования энергии. Энергия проявляет себя через действие сил. Поэтому, прибегая к коэффициенту полезного действия, мы можем лишь в общем плане установить проявление энергии через силы. Конкретные этапы силового взаимодействия остаются для нас неясными. В других случаях КПД просто неприемлем для технических характеристик. Например, при движении корабля возникает энергия в потоках взаимодействия и энергия топлива, используемая им в силовой установке типа двигателя внутреннего сгорания. С помощью только КПД мы не сможем установить связь между двумя этими энергиями. Следовательно, мы здесь не сможем применить закон сохранения энергии. Коэффициент эффективного использования сил даёт возможность связать между собой эти виды энергии. КЭИС приемлем также для характеристики любой другой техники и любого другого силового взаимодействия. В таких случаях он становится очень важным показателем. Например, современная техника характеризуется многоплановым силовым взаимодействием. Применив КЭИС, мы для каждого участка количественно выявим показатель этого силового взаимодействия. Тогда мы сможем охарактеризовать каждый участок или каждый силовой узел любой машины и сказать, какой из этих узлов требует дальнейшего совершенствования, а какой не требует. Ведь КЭИС даёт нам предельные показатели любого силового взаимодействия. (С)

**СОДЕРЖАНИЕ ПРИЛОЖЕНИЯ:**

**ПРИМЕР №1**

Требуется определить состояние жидкостей и газов в различных резервуарах и объёмах.

Здесь имеется в виду такое состояние жидкостей и газов, когда практически говорят, что они находятся под давлением. С точки зрения механики жидкости и газа, это состояние определяется как состояние покоя жидкостей и газов в скалярном силовом поле. Это значит, что жидкости и газы заполняют полностью весь объём резервуара и что давления в любых точках этого объёма одинаковы и равны по величине.

Статические силы давления в этих резервуарах определяются в зависимости от состояния массы жидкости или газа. Непосредственно для жидкостей силы давления определяются в соответствии с законом Паскаля, то есть, например, жидкости в резервуар нагнетаются под определенным давлением. Для газов силы давления можно определить, помимо закона Паскаля, еще по уравнению состояния Клапейрона:

$$Pv = RT.$$

Потенциальная энергия этих объёмов определяется как произведение сил давления на объём. Если отнести энергию к единице объёма, то количество единиц сил давления будет равно количеству единиц потенциальной энергии. Это значит, что стрелка манометра остановится на одном и том же делении шкалы как при замере сил давления, так и при замере потенциальной энергии. При одинаковой величине потенциальной энергии жидкости и газа в случае преобразования энергии жидкости в работу мы не получим почти никакой работы. Для газа мы получим работу, величина которой определяется по уравнению энергии сжатия (VII.7). Как для установившегося вида движения, так и для состояния покоя, уравнение энергии механики жидкости и газа и показания манометра не учитывают энергию сжатия. Следовательно, это характерная особенность уравнений энергии механики жидкости и газа.

В связи с тем, что мы рассматриваем состояние реальных жидкостей и газов, то мы должны учесть вязкость и термодинамическое состояние. Вязкость в данном случае не учитывается, так как отсутствуют динамические силы давления. Термодинамические условия притока или убыли тепловой энергии определяются с помощью уравнения энергии сжатия (VII-7). Тогда действительная величина потенциальной энергии состояния покоя жидкостей и газов $U_д$ запишется таким равенством:

$$U_д = U \pm \int_1^{V_2} PdV\,, \tag{1}$$

где $U_д$ – действительная энергия объёма жидкости или газа, $U$ – первоначальная энергия объёма жидкости или газа, $\int_1^{V_2} PdV$ – потеря или прибыль энергии, связанная с термодинамическими условиями.

При вычислении энергии сжатия для записи термодинамических условий пользуются зависимостями термодинамических процессов. Отметим, что для жидкостей в таких случаях не вычисляют энергию сжатия, а пользуются коэффициентом линейного расширения, который по сути своей не отражает истинной количественной зависимости. Поэтому он затрудняет и усложняет подобные практические расчёты. Следовательно, для жидкостей должны быть получены зависимости типа зависимостей термодинамических процессов газа, чтобы для жидкостей можно было пользоваться уравнением энергии сжатия (VII-7).

Мы записали состояние жидкостей и газов для определённых замкнутых объёмов, которые определяются статическими силами давления и потенциальной энергией. Резервуар, как конкретное условие при практическом использовании жидкостей и газов, проявляет себя как замкнутый объём известной величины и как средство внешнего воздействия на жидкости и газы.

Отметим, что уравнение состояния газов Клапейрона с точки зрения механики жидкости и газа является уравнением энергии единицы веса газа, так как удельный объём есть объём единицы веса газа, а произведение этого объёма на давление выражает количественно энергию.

**ПРИМЕР №2**

Требуется определить взаимодействие жидкостей и газов, находящихся в состоянии покоя под действием векторного силового поля, с поверхностью твёрдого тела.

За векторное силовое поле в данном примере принимаем поле земного тяготения, под действием которого находятся жидкости. В таком состоянии они содержатся либо в искусственных, либо естественных водоёмах. Твёрдые предметы также могут находиться либо в толще воды, либо плавать на её поверхности. Во всех случаях их механическое взаимодействие с жидкостью происходит непосредственно через поверхность твёрдых тел, которая непосредственно соприкасается с жидкостью. Эта поверхность является не только поверхностью взаимодействия, но и границей раздела двух механик: механики твёрдого тела и механики жидкости и газа.

Механическое взаимодействие будет заключаться в том, что все твёрдые тела в данном примере будут находиться под силовым воздействием жидкости на поверхности соприкосновения этих тел с жидкостью.

В данном примере нам необходимо определить зависимостями это силовое взаимодействие. Покажем на рисунке 28 часть воображаемого бассейна. На этом рисунке показано, что поверхностью соприкосновения жидкости и твёрдого тела служит днище и стенка до высоты $h$ воображаемого бассейна. На нем также показаны две системы отсчёта: для механики жидкости и газа ею служит поверхность жидкости, относительно которой определяется высота $h$. Для механики твёрдого тела ею является декартова система координат. Мы разместили её так, чтобы одна из осей совпадала с направлением высоты, а другая бы лежала в плоскости поверхности жидкости.

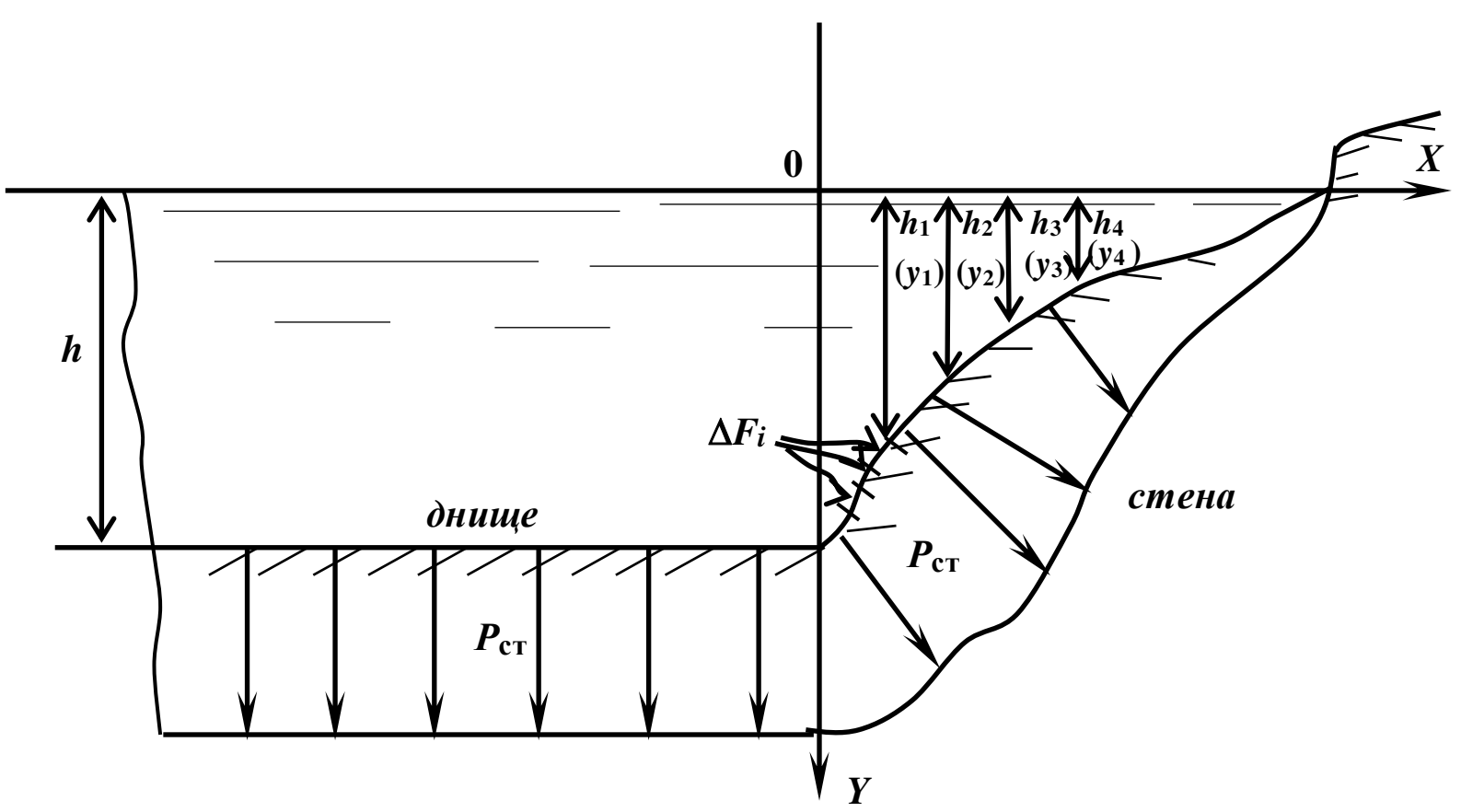

Рис. 28

В соответствии с положением о точке среды силы давления жидкости будут направлены в любой точке поверхности соприкосновения перпендикулярно к этой поверхности. Так как силы давления являются скалярными величинами, то направление их действия определяется поверхностью или плоскостью. Величина сил давления определяется по уравнению (II-15) в зависимости от высоты $h$. В свою очередь, действие сил давления жидкости уравновешивается на поверхности соприкосновения силовыми напряжениями, которые возникают в твёрдых телах. По отношению к твёрдому телу давление жидкости определяется как распределенные по поверхности силы. Распределение и сложение их в твёрдом теле определяется положениями механики твёрдого тела, которая рассматривает силы как векторные величины.

Дальше мы определяем распределение сил давления в бассейне и составляем условие равновесия. Силы, действующие со стороны твёрдого тела, будем определять как силу $R$, которая является функцией напряжений $\delta$, то есть $Rf(\delta)$.

Теперь рассмотрим действие сил давления жидкости на днище и стенки бассейна. Принимаем, что днище расположено параллельно нулевой поверхности жидкости, тогда величина сил, действующих на днище, будет равна произведению площади днища $F$ и величины сил давления, которые определяются по уравнению (II-15). Эта сила будет уравновешиваться силой, действующей со стороны твёрдого тела на поверхность днища. Запишем это условие:

$$Rf(\delta) = F\rho h w^2. \tag{2}$$

Вспомним, что в поле земного тяготения квадрат скорости $w^2$ равен по абсолютной величине ускорению $g$ в поле земного тяготения. Уравнением (2) мы записали условие равновесия для днища бассейна.

Принимаем, что профиль стенки бассейна по отношению к изменению высоты $h$ может быть любым. Поэтому для определения сил, действующих на стенку, воспользуемся приближенным методом. Для чего разобьём площадь стенки на ряд элементарных площадей, где можно приближенно считать, что высота $h_j$ для всей элементарной площадки одинакова. Тогда для каждой элементарной площадки можно записать условие равновесия в таком виде:

$$\Delta R_i f(\delta) = \Delta F_i \rho w^2 h_j\,. \tag{3}$$

Для всей поверхности бассейна условие равновесия запишется как сумма сил, действующих на всех элементарных площадках:

$$\sum_{i=1}^{n} R_i f(\delta) = \sum_{\substack{i=1 \\ j=1}}^{n} \Delta F_i \rho w^2 h_j\,. \tag{4}$$

Эта сумма определит нам условие равновесия по отношению к неинерциальной системе координат, то есть по отношению к действующим силам давления жидкости. По отношению к твёрдому телу и инерциальной системе координат сумму (4) надо записать с учётом направления действия сил давления и их равнодействующей, то есть как сумму проекций на оси инерциальной системы координат. В общем, действия над ними производятся в соответствии с положениями механики твёрдого тела. Таким образом, осуществляется переход и связь между положениями механики твёрдого тела и механики жидкости и газа.

Поверхность соприкосновения жидкости с твёрдым телом, как одно из условий практического использования жидкостей и газов, определяет в данном случае границу между положениями механики твёрдого тела и механики жидкости и газа.

**ПРИМЕР №3**

Требуется определить движение установившегося потока жидкости или газа в трубопроводах.

Для установившегося вида движения жидкостей и газов получены все необходимые зависимости. Согласно этим зависимостям, площадь сечения потока по участкам движения может быть различной. Они же определяют минимально возможную площадь сечения потока. В то же время из практики мы знаем, что для сохранения установившегося потока необходимо осуществлять плавный переход потока между его участками с различной площадью сечения. При наличии в потоке резких ступенчатых переходов от одной площади сечения к другой происходит значительное падение энергии потока, связанное с его геометрическими характеристиками. То есть эти потери непосредственно связаны с геометрической формой трубопровода, т.к. в данном примере его геометрические характеристики определяют геометрические характеристики потока. Целью данного примера является определение необходимых зависимостей и положений для участков плавного перехода потока с одной площади сечения на другую.

Дальше принимаем, что сечением трубопровода является прямоугольник, что переход потока от одной площади сечения к другой может осуществляться двумя способами: непосредственным изменением стенок трубопровода и размещением в трубопроводе твёрдого тела определённой формы. На рисунке 29 изобразим поток с сечением, изменённым в соответствии с двумя этими способами. Потоки, показанные на рисунках 29, *а* и 29, *б*, по расходу массы и по площадям сечения одинаковы. Установившийся поток требует симметрии для площадей своего сечения. По этой причине переход по сечению потока осуществляется за счёт изменения формы его нижней и верхней сторон. Отклонение от требований симметрии приводит к определённым потерям энергии. Теперь переходим к непосредственному рассмотрению участков потока в порядке движения жидкости в потоке.

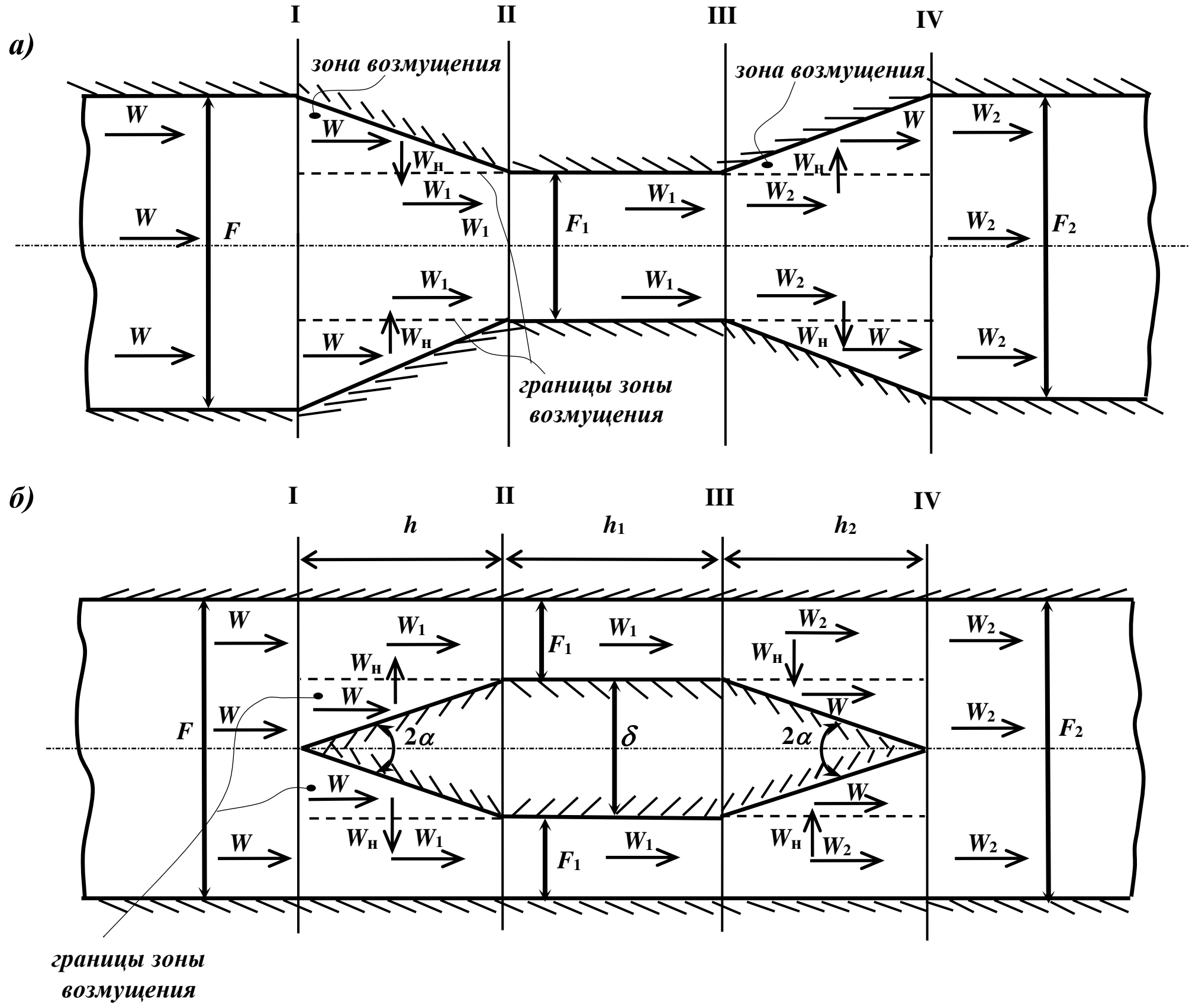

*Рис. 29*

***Участок I -II***

На этом участке происходит сужение потока. Такое изменение относится к пространственному изменению. Так как зависимости для установившегося движения не зависят от параметров пространства и времени, то для данного участка мы не можем воспользоваться ими. Поэтому нам придётся взять такие зависимости, которые связаны с пространственным изменением потока. Необходимые зависимости относятся к плоскому установившемуся виду движения.

Воспользуемся основным положением этого вида движения, которое утверждает, что пространственное движение потока происходит только в двух взаимно перпендикулярных направлениях. Одно направление нам известно – оно совпадает с направлением движения установившегося потока, то есть с направлением линии тока. Следовательно, второе направление движение будет иметь перпендикулярное к ней направление, то есть оно будет перпендикулярно оси потока. Теперь мы можем записать движение жидкости на участке потока I-II с помощью уравнений движения и сил. Дальнейшие рассуждения будем вести относительно рисунка 29, *б*, т.к. он образует одинаковую картину движения с рисунком 29, *а*, но более удобен для наглядного объяснения.

***Уравнения движения зоны возмущения***

Границы зоны возмущения участка потока I-II определяются длиной участка $h$, толщиной центрального тела $\delta$, шириной потока $Ш$. Дополнительной характеристикой является угол $\alpha$ (рис. 29, *б*).

Объём участка I-II, который расположен вне зоны возмущения, назовём просто объёмом потока участка I-II. Мы определили геометрические характеристики зоны возмущения.

Через плоскость I-I в зону возмущения будет поступать жидкость со скоростью $W$ и плотностью $\rho$. Плотность и скорость соответствуют величинам, которые определяются зависимостями предыдущего участка. Тогда уравнение движения для плоскости I-I можно будет записать в таком виде:

$$M_п = \rho W \cdot \delta Ш \qquad (5)$$

Уравнение (5) является уравнением движения зоны возмущения в плоскости I-I. С помощью уравнения (5) определяется поступательный расход массы в единицу времени $M_{п}$ в зоне возмущения. Этот расход вытеснит из зоны возмущения через ее границы в объём потока нормальный расход массы в единицу времени $M_{н}$. Нормальный и поступательный расходы массы зоны возмущения равны друг другу по абсолютной величине. В то же время нормальный расход массы будет отрицательным, т.к. он вытекает из зоны возмущения. Тогда уравнение для нормального расхода массы будет иметь вид:

$$M_{н} = -\rho W_{н} \cdot 2hШ \qquad (6)$$

Уравнение (6) является уравнением движения нормального движения в зоне возмущения. Из этого уравнения мы можем определить нормальную скорость движения $W_{н}$ зоны возмущения.

Расход массы в единицу времени в объёме потока участка I-II будет равен расходу массы всего установившегося потока $M$, а площадь сечения равна площади сечения участка II-III. Следовательно, уравнение движения этого объёма будет одинаковым с уравнением движения участка II-III. Запишем его:

$$M = \rho W_1 F_1, \qquad (7)$$

где $W_1$ – скорость участка потока II-III, $F_1$ – площадь сечения потока на участке II-III.

Мы получили все необходимые уравнения движения участка потока I-II, которых три – (5), (6) и (7).

***Уравнения сил участка потока I-II***

Для плоскости I-I зоны возмущения уравнение сил будет иметь вид:

$$P_{пр.п} = P_{ст.п} + \rho W^2. \qquad (8)$$

Для нормального движения на границе зоны возмущения оно будет иметь вид:

$$P_{пр.н} = P_{ст.н} - \rho W_{н}^2. \qquad (9)$$

Для объёма потока участка I-II оно будет иметь вид:

$$P_{пр} = P_{ст1} + \rho W_1^2. \qquad (10)$$

Уравнения (8), (9) и (10) являются уравнениями сил участка потока I-II.

Величины сил давления в этих уравнениях вычисляются почленно. Сначала вычисляют динамические силы давления, так как плотность и скорость для каждого из них известны. Они были получены по уравнениям движения. Дальше, исходя из условия постоянства полной энергии потока для любой точки объёма участка I-II, определим величины статических сил давления в уравнениях (8), (9) и (10). Затем мы получим принятые, или суммарные, силы давления $P_{пр}$ для каждого из них. Решив уравнения движения и сил участка I-II для конкретного установившегося потока жидкости, мы получим определённые количественные величины для него.

Теперь перейдём к исследованию, которое будет заключаться в том, что мы рассмотрим влияние изменения угла $\alpha$ на характеристики нормального движения зоны возмущения и всего движения в целом на участке потока I-II. Принимаем диапазон изменения угла $\alpha$ от 0° до 90°.

При стремлении угла $\alpha$ к нулю длина $h$ участка потока I-II будет стремиться к бесконечности. Это значит, что в уравнении движения (6) для нормального расхода массы в зоне возмущения площадь нормального движения будет стремиться к бесконечности. В связи с тем, что расход массы является постоянной величиной, то нормальная скорость движения $W_{н}$ будет стремиться к нулю.

Теперь полагаем, что угол $\alpha$ стремится к 90°. Тогда длина участка I-II $h$ и площадь нормального расхода в уравнении (6) будут стремиться к нулю. В связи с тем, что расход массы остаётся постоянной величиной, то нормальные скорости движения $W_н$ будут стремиться к бесконечности.

При изменении угла $\alpha$ в диапазоне исследования другие характеристики потока участка I-II останутся без изменения, то есть уравнения (5), (7) и (8), (10) и их переменные останутся неизменными.

Данное исследование интересовало нас по такому вопросу: каковы могут быть максимальные и минимальные силы давления, действующие на площадь сечения центрального тела ($\delta \cdot Ш$) на участке потока I-II. Величина этих сил давления определяется уравнениями сил (8) и (9) зоны возмущения этого участка.

При стремлении угла $\alpha$ к нулю нормальные динамические силы давления в уравнении (9) тоже стремятся к нулю. Это значит, что величина силы, действующей на площадь сечения центрального тела, будет определяться только уравнением (8), то есть поступательными динамическими и статическими силами давления. В этом случае мы будем иметь максимальные силы давления на площади сечения центрального тела.

При стремлении угла $\alpha$ к 90° нормальные динамические силы давления стремятся к бесконечности. Конечно, они не достигают этой величины, так как по условиям движения, которые были рассмотрены выше, динамические силы давления достигают своей максимальной величины, которая равна статическому давлению сил для полной энергии установившегося потока. Если угол $\alpha$ находится в близких пределах к максимальным динамическим силам давления, то в этом случае мы получим минимальные давления на площади сечения центрального тела.

Уменьшение сил давления здесь происходит за счёт уменьшения статических сил давления в уравнении (9). Так как нормальные динамические силы давления действуют перпендикулярно направлению потока, то они не оказывают никакого прямого воздействия на центральное тело. Они действуют на него как бы косвенным путем. Их увеличение приводит к уменьшению статического давления в уравнении (9). В то же время величина этих статических сил давления не может быть больше величины статических сил давления, определённых по уравнению (8), что, в свою очередь, приводит к уменьшению сил, действующих на площадь сечения центрального тела, конечно, за счёт уменьшения статических сил давления. Количественно это можно определить по распределению полной энергии потока между потенциальной и кинетической. Б*о*льшие нормальные скорости, по сравнению с поступательными скоростями движения, приводят к уменьшению статических сил давления.

Отсюда следует, что минимальные силы давления действуют на поперечное сечение центрального тела при максимальных нормальных скоростях движения в зоне возмущения участка потока I-II.

Следующее условие, которое нас тоже интересует, это обтекаемость, то есть движение потока на участке I-II без потерь полной энергии потока. Рассмотрим обтекаемость центрального тела в зависимости от его формы на участке потока I-II.

При стремлении угла $\alpha$ к нулю увеличивается длина профиля, но это не приводит к нарушениям в зависимостях сил, движения и энергии. Следовательно, никаких потерь полной энергии потока здесь не будет, то есть профиль центрального тела на участке потока I-II будет обтекаем.

При стремлении угла $\alpha$ к 90° профиль центрального тела на участке потока I-II стремится к плоскости II-II, а нормальные скорости – к бесконечности. Здесь мы наблюдаем нарушение распределения полной энергии потока между потенциальной и кинетической. В этом случае реальные максимально возможные нормальные скорости поддерживаются частично за счёт потерь полной энергии потока. Тогда подобный профиль относится к разряду плохо обтекаемых, так как он требует потерь полной энергии потока. Чтобы сделать профиль обтекаемым, выбирают угол $\alpha$ таким, чтобы он не нарушал условие распределения полной энергии потока между потенциальной и кинетической. Практика показывает, что обтекаемый профиль с окружностью вместо треугольника соответствует профилю с максимально возможным углом $\alpha$.

Мы рассмотрели условие обтекаемости профиля. Будем считать, что на этом задачу по участку потока I-II мы выполнили.

Участок потока II-III есть обыкновенный участок установившегося потока. Поэтому для него пригодны все зависимости установившегося вида движения. В связи с этим мы не будем его рассматривать.

***Рассмотрим движение на участке потока III - IV.***

Сначала сопоставим картину движения на участках потока I-II и III-IV. Принципиальное различие здесь будет заключаться в том, что скорости движения в зоне возмущения этих участков будут иметь противоположные направления. Характер изменения движения на участке III-IV происходит относительно последующего участка потока. Поэтому непосредственно в объёме потока участка III-IV скорость будет равна скорости на последующем участке $W_2$. Величина расхода массы в единицу времени зоны возмущения определяется в плоскости IV-IV для площади сечения зоны возмущения по уравнению движения. Запишем его:

$$M_п = -\rho W \cdot \delta Ш \tag{11}$$

Для нормального движения на границе зоны возмущения уравнение движения будет иметь такой вид:

$$M_н = \rho\, W_н \cdot h_2\ Ш. \tag{12}$$

Расходы масс в уравнениях (11) и (12) равны друг другу, то есть $M_п = M_н$. Скорость движения в объёме потока участка III-IV определяется по уравнению движения следующего участка, поэтому

$$M = \rho W_2 F_1. \tag{13}$$

Мы получили все необходимые уравнения движения для участка потока III-IV.

Запишем теперь уравнения сил для этого участка.

Уравнение сил поступательного движения зоны возмущения участка III-IV будет иметь вид:

$$P_{пр.п} = P_{ст.п} - \rho W^2. \tag{14}$$

Уравнение сил нормального движения зоны возмущения участка III-IV будет иметь вид:

$$P_{пр.н} = P_{ст.н} + \rho W_н^2. \tag{15}$$

Уравнение сил объёма потока участка III-IV будет иметь вид:

$$P_{пр2} = P_{ст2} + \rho W_2^2. \tag{16}$$

Мы получили уравнения сил для участка потока III-IV. Они тоже вычисляются почленно, как и для участка I-II.

При изменении угла $\alpha$ от 0° до 90° профиля центрального тела будут изменяться нормальные скорости движения $W_н$ зоны возмущения и связанные с нею характеристики. Эти изменения угла $\alpha$ центрального тела тоже связаны с исследованием величины сил, действующих на площадь сечения центрального тела ($\delta \cdot Ш$), и исследованием обтекаемости тела.

В зоне возмущения участка потока III-IV на профиль центрального тела будут действовать статические и динамические силы давления поступательного и нормального движения. Рассмотрим их действие.

Поступательные динамические силы давления направлены от поверхности центрального тела, поэтому они не оказывают на него силового воздействия.

Нормальные динамические силы давления направлены перпендикулярно оси потока. Тем самым они уравновешивают друг друга и не оказывают осевого воздействия на центральное тело.

Статические силы давления оказывают силовое воздействие на профиль центрального тела. Их величина зависит от величины динамических сил давления.

Изменение угла $\alpha$ профиля центрального тела изменяет величину нормальных динамических сил давления, которые, в свою очередь, влияют на величину статических сил давления.

При угле $\alpha$, стремящемся к нулю, нормальные динамические силы давления тоже стремятся к нулю. В этом случае величина статических сил давления определяется только поступательными динамическими силами давления зоны возмущения. Поэтому по отношению к нормальным динамическим силам давления статические силы давления будут иметь максимальную величину.

При угле $\alpha$, стремящемся к 90°, нормальные динамические силы будут стремиться к бесконечности. В этом случае величина статических сил давления будет зависеть от величины нормальных динамических сил давления зоны возмущения участка потока III-IV. В зависимости от этих сил мы здесь получим минимальное значение статических сил давления. Следовательно, варьируя углом $\alpha$ профиля центрального тела на участке потока III-IV, мы изменяем в определённом диапазоне величину нормальных динамических сил давления и величину статических сил давления.

Обтекаемость профиля центрального тела на участке потока III-IV нарушается при стремлении угла $\alpha$ к 90°. В этом случае мы можем нарушить распределение полной энергии потока между потенциальной и кинетической. Максимальные нормальные скорости зоны возмущения в этом случае будут поддерживаться за счёт потерь полной энергии потока.[6]

**ПРИМЕР № 4**

В данном примере требуется определить силы, действующие на крыло, которое перемещается в среде неподвижной жидкости или газа.[7]

Под термином «крыло» принято понимать твёрдое тело определённого асимметричного профиля, которое при движении в среде неподвижного газа или жидкости создаёт определённую подъёмную силу. Среда здесь определяется как некоторый объём жидкости или газа, который находится в состоянии покоя под действием векторного или скалярного силового поля. Следовательно, жидкости или газы в этом случае обладают определённым запасом потенциальной энергии и имеют определённые статические силы давления.

Дальше полагаем, что крыло движется с определённой постоянной скоростью в среде жидкости или газа. Покажем его движение на рисунке 30. Передний профиль крыла образован окружностью с радиусом $r$, задний – треугольником с углом задней кромки крыла равной углу $\alpha$.

Крыло движется с постоянной скоростью $W$. В свою очередь, жидкость или газ будут набегать на крыло тоже с постоянной скоростью равной скорости движения этого крыла. В тоже время жидкости и газы при движении относительно профиля крыла испытывают пространственное изменение. Поэтому для решения этой задачи воспользуемся зависимостями установившегося и плоского установившегося видов движения, как это сделано в примере № 3. В соответствии с этим примером определим границы зон возмущения для профиля крыла, как показано на рисунке 30. Затем переходим к определению зависимостей для данного профиля.

1. За полную энергию набегающего установившегося потока принимаем удельную энергию среды $U_{\text{ср}}$.
2. Составляем уравнение движения зоны возмущения, которая ограничивается плоскостями I и II. В плоскости I для поступательной плоскости исследования оно будет иметь вид:

$$M_{\text{п}1} = \rho W r. \tag{17}$$

В уравнении (17) для площади сечения мы принимаем длину крыла за единицу длины. Поэтому площадь здесь записана произведением единицы длины на радиус $r$. Для других зависимостей мы тоже будем относить площадь к единице длины.

[6] Вопрос обтекаемости твёрдых тел рассмотрен в работе «Движение твёрдых тел в жидкостях и газах с точки зрения механики безынертной массы». В данном примере нет ещё понятие о тыльном сопротивлении.

[7] См. работу «Движение твёрдых тел в жидкостях и газах с точки зрения механики безынертной массы».

В уравнении (17) нам известна плотность $\rho$. Как для жидкостей, так и для газов она равна плотности неподвижной среды. Также известна скорость $W$ – она равна скорости движения крыла, но имеет противоположное направление. Следовательно, по уравнению движения (17) мы можем определить расход массы в единицу времени для зоны возмущения I – II, так как этот расход будет равен нормальному расходу массы зоны возмущения, то есть $M_{\text{п}1} = M_{\text{н}1}$.

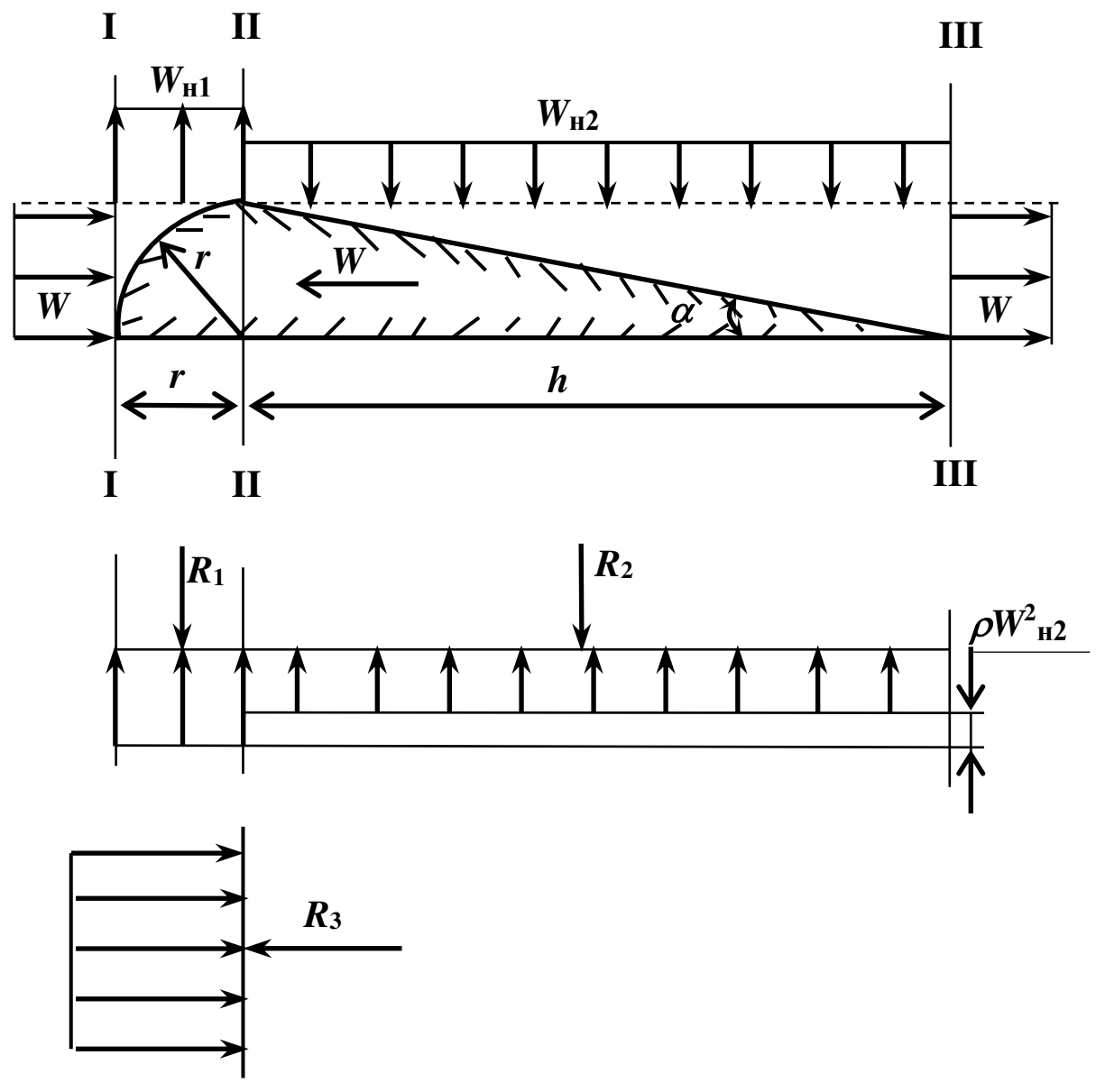

Рис. 30

Запишем уравнение движения для нормального движения на площади границы зоны возмущения. Оно будет иметь вид:

$$M_{\text{н}1}=\rho W_{\text{н}1} r. \tag{18}$$

Из уравнения движения (18) мы можем определить нормальную скорость движения зоны возмущения I – II, так как в этом уравнении нам известны все величины, кроме нормальной скорости. В данном конкретном случае она будет равна поступательной скорости зоны возмущения, в связи с тем, что площади нормального и поступательного потоков зоны возмущения равны между собой.

Уравнение движения для объёма потока участка I – II составлять не требуется. Так как среда неподвижна, и объём её есть бесконечно большой.

Мы получили полный комплект уравнений движения для зоны возмущения участка профиля крыла I - II.

3. Составляем уравнения сил для зоны возмущения участка I – II.

Для поступательного движения зоны возмущения участка I – II оно будет иметь вид:

$$P_{\text{пр.п}} = P_{\text{ст.п}} + \rho W^2. \tag{19}$$

Для нормального движения зоны возмущения участка I – II оно будет иметь вид:

$$P_{\text{пр.н}} = P_{\text{ст.н}} - \rho W_{\text{н}\,1}^2 \tag{20}$$

4. Затем по уравнению энергии установившегося вида движения или по уравнению Бернулли определяем величину статических сил давления для нормального и поступательного потоков участка I – II. Полная энергия потока нам известна – это потенциальная энергия неподвижной среды. Кинетическую энергию мы можем вычислить, так как нам известна плотность и скорость нормального и поступательного потоков зоны возмущения. В нашем конкретном примере нормальные и поступательные статические силы давления будут равны друг другу. Вычислив статические силы давления, мы сможем определить все действующие силы на участке I – II.

Теперь нам остаётся определить силы, действующие на профиль крыла участка II – III.

1. Полная энергия установившегося потока будет равна потенциальной энергии среды $U_{ср}$.
2. Уравнение движения поступательного движения плоскости III для площади зоны возмущения участка II – III будет иметь вид:

$$M_{п1}=\rho Wr. \qquad (21)$$

Это уравнение по всем величинам равно уравнению (17).

*Уравнение движения для нормального движения* на границе зоны возмущения участка II – III будет иметь вид:

$$M_{н2}=\rho W_{н2}h. \qquad (22)$$

Из этого уравнения мы сможем определить величину нормальной скорости движения в зоне возмущения.

3. Уравнение сил поступательного движения зоны возмущения участка II – III будет иметь вид:

$$P_{пр.п} = P_{ст.п} - \rho W^2. \qquad (23)$$

Уравнение сил нормального движения зоны возмущения участка II – III будет иметь вид:

$$P_{пр.н} = P_{ст.н} - \rho W_{н2}^2. \qquad (24)$$

Затем по уравнению энергии установившегося вида движения делаем распределение полной энергии потока на потенциальную и кинетическую части для поступательного и нормального движения и определяем статические силы давлений этих движений на участке II – III.

Вычислив статические силы давления зоны возмущения участка II – III, мы получим все количественные величины для всего профиля крыла. Теперь мы сможем составить условие равновесия для движущегося крыла.

### ***Подъёмная сила***

Подъёмные силы крыла действуют перпендикулярно направлению его движения. На участке потока I – II со стороны зоны возмущения действуют статические и динамические силы давления. Нормальные динамические силы давления этого участка направлены от профиля крыла. Поэтому они не оказывают на него воздействия. Поступательные динамические силы давления этого участка действуют на профиль крыла в направлении его движения. Поэтому они тоже не влияют на подъёмную силу профиля.

Со стороны зоны возмущения участка I – II на профиль крыла будут действовать только статические силы давления, которые для поступательного и нормального движения в нашем примере одинаковы.

Со стороны невозмущённой среды на площадь участка I – II профиля крыла будут действовать статические силы давления, которые соответствуют полной энергии установившегося потока, или потенциальной энергии среды. По величине они больше статических сил давления зоны возмущения участка I – II. Поэтому за счёт разности этих давлений образуется подъёмная сила крыла на участке I – II. Если обозначить подъёмную силу крыла участка I – II через $R_1$, то она будет равна:

$$R_1 = (P_{ст.ср} - P_{ст1})\cdot r. \qquad (25)$$

В зоне участка II – III на профиль крыла тоже будут действовать статические и динамические силы давления.

Со стороны зоны возмущения участка II – III действует поступательные динамические силы давления, которые действуют в противоположную скорости движения крыла сторону, и действие их направлено от профиля. Поэтому они не оказывают силового воздействия на него.

Нормальные динамические силы давления участка II – III действуют непосредственно на профиль крыла. Их направление совпадает с направлением действия подъёмной силы крыла.

Статические силы участка II – III в зоне возмущения тоже действуют на профиль крыла. Их действие мы тоже должны учесть при определении подъёмной силы крыла. Мы знаем, что, в зависимости от величины нормальных и поступательных сил давления зоны возмущения, статические силы могут быть разными. В данном случае статическое давление принимается непосредственно в зависимости от большей величины нормальных или поступательных динамических сил давления. В нашем случае поступательные динамические силы имеют б*о*льшую величину, чем нормальные динамические силы давления. Поэтому для расчёта подъёмной силы крыла мы возьмём статические силы давления поступательного движения зоны возмущения участка II – III.

Со стороны невозмущённого потока будут действовать статические силы давления, величина которых соответствует величине полной энергии установившегося потока или потенциальной энергии среды.

Обозначим подъёмную силу участка профиля через $R_2$, тогда её величина будет равна разности статического давления невозмущённой среды с нормальными динамическими силами возмущённого участка II – III и поступательными статическими силами того же участка профиля. Запишем величину подъёмной силы:

$$R_2 = (P_{\text{ст.ср}} - \rho W_{\text{н2}}^2 - P_{\text{ст.п}}) \cdot h. \qquad (26)$$

Мы получили подъёмную силу для всего профиля крыла, изображенного на рисунке 30.

Зная, какие характеристики влияют на величину подъёмной силы крыла, можно сравнительно просто управлять ею.

### *Лобовое сопротивление крыла*

Нам остаётся определить лобовое сопротивление крыла. Опять начнём с распределения сил давления на участке профиля I – II. Силы, определяющие лобовое сопротивление крыла, действуют в направлении его движения. Такими силами в зоне возмущения профиля участка I – II являются поступательные динамические и статические силы давления. В зоне возмущения участка II – III поступательные динамические силы давления направлены от профиля крыла. Поэтому они не оказывают на него силового воздействия, и мы их не будем учитывать при определении лобового сопротивления. Остаются только поступательные статические силы давления этого участка. Обозначим силу сопротивления через $R_3$. Она будет равна сумме поступательных динамических и статических сил давления за вычетом поступательных сил давления зоны возмущения участка профиля II – III. Запишем эту силу лобового сопротивления:

$$R_3 = (P_{\text{ст.п1}} + \rho W^2 - P_{\text{ст.п2}}) \cdot r. \qquad (27)$$

В нашем примере поступательные статические силы участка профиля I – II и II – III равны между собой. Тогда сила лобового сопротивления будет равна только поступательным динамическим силам давления зоны возмущения участка I – II, то есть:

$$R_3 = \rho W_{\text{п1}}^2 \cdot r. \qquad (28)$$

Мы определили подъёмную силу и лобовое сопротивление крыла в зависимости от его профиля.

Отметим, что при расчёте подъёмной силы крыла в реальных вязких жидкостях и газах необходимо будет нормальные динамические силы давления в уравнении (26) умножить на коэффициент кинематической вязкости $\mu$, а при расчёте сил лобового сопротивления необходимо будет умножить поступательные силы давления в уравнениях (27) и (28) на динамический коэффициент вязкости $\nu$.

В примере № 4 мы рассмотрели влияние условий практического использования жидкостей и газов в их качестве неподвижной среды при движении профиля крыла. Как видим, и эти условия тоже требуют решения своих проблем.

**ПРИМЕР № 5**

В данном примере нам необходимо определить состояние набегающего потока жидкости или газа по энергетическому состоянию при движении твёрдого тела в неподвижной среде. При условии, что твёрдое тело в каждый новый рассматриваемый момент времени будет двигаться с большей скоростью, чем в предыдущий момент движения.

В общем, сущность этого примера будет состоять в том, чтобы определить изменение энергетического состояния набегающего потока в зависимости от скорости твёрдого тела.

Потенциальная механическая энергия естественной среды, то есть воды или воздуха, определяется природными условиями, которые создаёт для них наша планета с помощью силовых полей. Величину её мы можем получить при помощи соответствующих замеров.

По отношению к движущемуся предмету среда образует поток установившегося вида движения. Потенциальная энергия среды является полной энергией этого потока. Движущийся предмет, согласно уравнению энергии установившегося вида движения, или уравнению Бернулли, производит перераспределение полной энергии потока или потенциальной энергии среды между потенциальной и кинетической энергией этого образовавшегося потока.

Из условий движения установившегося потока реальных невязких жидкостей и газов в трубопроводах нам известны предельные условия распределения полной энергии потока между кинетической и потенциальной. Они выражаются в том, что кинетическая и потенциальная энергия потока по величине достигают, каждая в отдельности, половины полной энергии потока ($U_{\text{к}} = U_{\text{п}} = \frac{U}{2}$). Тогда скорости движения жидкостей и газов становятся определённой постоянной величиной, которую можно увеличить лишь при увеличении полной энергии установившегося потока. Для газа скорость движения при предельных условиях распределения достигает скорости звука и называется критической скоростью.

В тоже время, мы знаем, что различные предметы могут двигаться в среде жидкостей и газов со скоростью, которая превышает, например, скорость звука для данной среды. По этой причине энергетическое состояние образующегося набегающего потока должно быть иным, чем это записано уравнением механической энергии установившегося вида движения или уравнением Бернулли. В этом примере требуется определить это новое энергетическое состояние.

Рассматриваем движение предмета со сверхзвуковой скоростью в среде газа. Что здесь нам известно из практики? Только величина скорости движения и локальное качественное изменение среды вокруг движущегося предмета, которое, например, выражается в повышенных температурах. Если мы теперь попытаемся записать для него уравнение распределения энергии в обычном виде, то нам этого не удастся сделать, поэтому мы должны найти иное распределение энергии набегающего потока.

Как мы выше установили, что при движении предмета в среде газа со сверхзвуковой скоростью образуется установившийся поток газа с механической и термодинамической энергией. В разделе о реальных жидкостях и газах мы установили, что полная энергия такого потока записывается уравнением энергии (VII-8). Выпишем его ещё раз:

$$U_{\text{д}} = U_{\text{п}} + \int_{0}^{P_2} dP \pm \int_{1}^{V_2} PdV. \qquad \text{(VII-8)}$$

Согласно этому уравнению и положениям, связанным с ним, для установившегося потока следует, что при изменении термодинамических свойств массы газа или жидкости механическая энергия потока остаётся неизменной, а прирост энергии происходит за счёт изменения термодинамических свойств массы жидкости и газа. В этом уравнении механическая и термодинамическая энергии являются отдельными самостоятельными членами. В этом случае распределение полной механической энергии между потенциальной и кинетической энергией достигает своего крайнего предела. То есть для невязких жидкостей и газов величины кинетической и потенциальной энергий потока становятся равны половине величины полной механической энергии потока. Из этого условия следует, что верхний предел определённого интеграла кинетической энергии (VII-8) достигает своей предельной величины и становится равным максимальному динамическому значению сил давления $P_{\max}$ . Мы знаем, как можно определить величину этого давления. Она равна величине статического давления для полной механической энергии потока. В нашем случае она равна статическому давлению для неподвижной среды с соответствующей величиной потенциальной энергии.

Член уравнения (VII-8), учитывающий изменение термодинамической энергии, будет в нашем случае положительной величиной, так как у нас происходит прирост энергии. Теперь перепишем уравнение (VII-8) с учётом этих изменений:

$$U_{\text{д}} = U_{\text{п}} + \int_{0}^{P_{\max}} dP + \int_{1}^{V_2} PdV. \qquad (29)$$

Мы получили уравнение энергии установившегося потока газа, образующегося при движении предмета со сверхзвуковой скоростью в среде газа.

Теперь нам необходимо выразить энергию через характеристики потока. Для чего сначала запишем кинетическую энергию потока через его характеристики. Тогда получим:

$$U_{\text{к max}} = \frac{\rho W^2}{2} = \text{const}. \qquad (30)$$

Рассмотрим все характеристики уравнения кинетической энергии (30). Максимальная кинетическая энергия $U_{\text{к .max}}$ есть величина постоянная в данном примере и равная половине потенциальной энергии среды или половине полной энергии установившегося потока ($U/2$). Следовательно, её величину мы можем определить.

Скорость потока в этом случае можно определить через скорость движения предмета в среде. Скорость движения предмета в среде и скорость движения газа в возмущённой зоне среды будут равны друг другу. Это определение относится, прежде всего, к поступательной скорости движения. Следовательно, скорость движения газа мы тоже можем определить.

В этом уравнении остаётся неизвестным для нас только плотность газа $\rho$. Используя уравнение (30), мы можем определить её величину в каждом конкретном случае. Проанализировав изменение плотности газа в зависимости от скорости, мы придем к выводу, что с увеличением скорости при сверхзвуковом движении предмета плотность среды уменьшается. При этом зависимость здесь квадратичная. Мы получили кинетическую энергию потока, выраженную через его характеристики.

Чтобы определить действительную энергию потока $U_{\text{д}}$, мы должны знать величину термодинамической энергии в уравнении (29). Её величину мы можем определить только по изменению плотности. Но для этого надо знать соответствующую зависимость термодинамического процесса. С этим вопросом обратимся к термодинамике. Согласно её положениям в данном случае мы должны были бы применить зависимость адиабатического процесса, но его зависимость неприменима для нашего случая. Так как с ростом энергии уменьшается плотность среды. Пожалуй, в данном случае нам подойдет зависимость изобарного процесса, но этот процесс рассматривает изменение термодинамических характеристик газа при постоянном давлении, а не при постоянной величине механической энергии. Так что и здесь мы можем попасть в ловушку. Следовательно, термодинамика не даёт чёткого ответа в этом вопросе. Поэтому для дальнейших рассуждений принимаем, что изменение плотности газа в потоке происходит по некоторой зависимости политропного процесса. Будем считать, что нам известна зависимость термодинамических изменений состояния массы газа.

Тогда величину термодинамической энергии мы определим следующим образом. Пределы изменения плотности газа мы определим по уравнению кинетической энергии (30). Затем с помощью уравнения термодинамического процесса найдем пределы изменения единицы а ($V_2$) для определённого интеграла уравнения (29), который выражает количество термодинамической энергии. После этого заменим в этом интеграле давление $P$ с помощью зависимости термодинамического процесса и вычислим полную величину термодинамической энергии установившегося потока газа. Следовательно, теперь мы можем определить количество термодинамической энергии для установившегося потока газа, образованного движущимся предметом, скорость которого превышает скорость звука среды неподвижного газа. Сложив потенциальную энергию среды и термодинамическую энергию, мы получим действительную энергию потока газа.

При движении предметов с ещё большими скоростями, то есть со скоростями, которые известны нам из практики, никаких других качественных изменений среды не наблюдалось. Поэтому не будем дальше рассматривать движение предметов в среде газа с ещё большими скоростями. Хотя скачок уплотнения оказался скачком «разуплотнения» при рассмотрении движения предмета в среде со сверхзвуковыми скоростями, но мы всё равно получили необходимые зависимости, которые характеризуют качественное изменение в газах в зависимости от их термодинамических свойств.[8] Следовательно, скачок уплотнения мы должны понимать как переход массы газа от одних термодинамических свойств к другим при определённых условиях движения.

Теперь рассмотрим движение предмета с постоянной скоростью в среде жидкостей. При движении предметов у поверхностей водных бассейнов нашей планеты до глубины не более 5 метров уже при скорости 10-15 м/сек происходит предельное распределение полной энергии набегающего потока или потенциальной энергии среды между потенциальной и кинетической, то есть, без учёта вязкости, величина потенциальной и кинетической энергии потока достигает половины полной энергии потока. Поэтому, при движении предмета в бассейнах нашей планеты со скоростями больше 15 м/сек, мы должны будем учитывать в уравнении полной энергии потока ещё и термодинамическую энергию. Следовательно, при скорости движения предмета 10-15 м/сек происходит скачок уплотнения, аналогичный скачку уплотнения газов. С этим скачком у людей не было никаких неприятностей. Пожалуй, по этой причине он остался незамеченным. В то же время при определении действительной энергии набегающего потока мы обязаны воспользоваться уравнением (29). Порядок определения этой энергии такой же, как и для газов. Полная механическая энергия набегающего потока берётся как потенциальная энергия среды. Плотность жидкости в возмущённой зоне определяется по уравнению (30). Термодинамическая энергия определяется по изменению плотности в соответствии с зависимостью термодинамического процесса. Затем определяется действительная энергия набегающего потока как сумма полной механической энергии набегающего потока и термодинамической энергии.

Отметим, что для жидкости в термодинамике нет каких-либо процессов, которые бы через зависимости выражали её термодинамические свойства. Так как жидкость считается несжимаемой. Просто переход тепловой энергии в механическую определяют при помощи соответствующего эквивалента. Будем надеяться, что она может быть «разжимаемой».

Считаем, что мы получили все необходимые зависимости для первого скачка термодинамических свойств жидкости.

С увеличением постоянной скорости движения предмета в среде жидкости растёт и температура жидкости в набегающем потоке. Наконец, скорость достигнет такой величины, когда термодинамическое состояние жидкости станет неустойчивым и жидкость начнёт переходить в пар. В обычной жизни это состояние называют кипением жидкости. Для жидкости набегающего потока подобное состояние будем называть кавитацией. Для неё это будет второй скачок изменения чисто термодинамических свойств. В водных бассейнах нашей планеты для набегающего потока кавитация наступает при движении предмета со скоростью порядка 40-45 м/сек.

Кавитацию можно назвать своеобразным процессом кипения жидкости в набегающем потоке, который образован движущимся в среде жидкости предметом. С помощью энергетических зависимостей можно определить только границы для каждой жидкости, в пределах которых существует кавитация.

Если дальше рассматривать увеличение постоянной скорости предмета, то мы увидим, что при достижении определённой величины скорости, кавитация в набегающем потоке прекращается и жидкость поступает в поток сразу в газообразном или парообразном состоянии. Дальнейшее увеличение скорости движущегося предмета надо рассматривать как увеличение его скорости при движении в среде газа.

---

[8] См. работу «Движение твёрдых тел в жидкостях и газах с точки зрения механики безынертной массы». Автор приходит к выводу, что плотность воздушной среды не изменяется при достижении сверхзвуковой скорости, нет ни уплотнения, ни разрежения. Происходит изохорный процесс. В нагретом газе скорость звука возрастает, поэтому само движущееся тело не достигает новой скорости звука.

Отметим, что движение предмета в жидкостях и газах рассматривалось как движение твёрдого тела абсолютно обтекаемой формы. Для необтекаемых тел, например, таких как плоскость, соответствующие скачки наступают при меньших скоростях движения, чем для обтекаемых.

Величину изменения плотности в набегающем потоке жидкости и газа можно контролировать путем измерения объёма или толщины пристеночного слоя, так как его величина определяет геометрические размеры набегающего потока.

Из этого примера следует, что набегающий установившийся поток жидкостей или газов воспринимает механическую энергию движущегося тела в виде распределения полной механической энергии среды между потенциальной и кинетической, а также за счёт изменения термодинамического состояния жидкостей и газов в этом потоке.

В примере № 5 условия практического использования жидкостей и газов определяют ряд проблем, которые касаются не только механики жидкости и газа, но и термодинамики[9].

**ПРИМЕР № 6**

Даны два установившихся потока жидкости или газа с полной энергией потока $U$ и с постоянным расходом массы в единицу времени $M$. Потоки вытекают в окружающую среду с потенциальной энергией $U_{ср}$. Требуется определить размеры и профиль рабочей полости турбины, которая используется для преобразования энергии этих потоков в механическую работу.

Турбины относятся к лопастным машинам. В соответствии с положениями механики жидкости и газа принцип их работы основан на плоском установившемся виде движения жидкостей и газов. В зависимости от характера движения жидкостей и газов в рабочей полости все турбины можно разделить на два типа: радиальные и осевые. В свою очередь рабочая полость этих турбин делится на два участка. Границы участков определяются в соответствии с энергетическим состоянием жидкостей и газов в объёме плоского установившегося потока. Энергетическое состояние жидкостей и газов показано на рис. 27, гл.V. Согласно этому рисунку первый участок располагается от точки $A_2$ до точки $A_0$. Для турбины он определяется как участок от входа до критического сечения. На этом участке происходит преобразование части потенциальной энергии потока в работу и увеличение скорости потока от входных скоростей до максимально возможных скоростей в критическом сечении.

Второй участок располагается от точки $A_0$ до точки $A_3$. Для турбины он определяется как участок от критического до выходного сечения. На этом участке происходит преобразование части кинетической энергии в работу. Если это газовая турбина, то на этом же участке дополнительно происходит преобразование энергии сжатия в работу. Одновременно происходит снижение скоростей потока от максимально возможных до выходных скоростей. По этому участку различаются газовые и жидкостные турбины.

Турбины, полость которых состоит из двух участков, дают возможность целиком использовать полную энергию установившегося потока жидкости или газа.

При практическом использовании радиальных турбин для конкретных условий применения не всегда возможно сделать оба её участка радиальными. В этом случае, например, можно будет сделать первый участок турбины радиальным, а второй – осевым. В нашем примере возьмём турбину именно такого типа, так как она даёт возможность рассмотреть одновременно конструктивные особенности радиальных и осевых турбин.

Зависимости, определяющие движение, силовое воздействие и распределение энергии в полости турбины, относятся к плоскому установившемуся виду движения идеальной жидкости. Для реальных жидкостей и газов их необходимо будет дополнить коэффициентами вязкости и зависимостями, связанными с сжимаемостью.

Теперь перейдем к определению рабочей полости турбины. Покажем её на рис. 31.

1.Определяем площади сечения турбины на входе (точка $A_2$), на выходе (точка $A_3$) и в критическом сечении (точка $A_0$). Для этого воспользуемся зависимостями установившегося вида

[9] См. работу «Движение твёрдых тел в жидкостях и газах с точки зрения механики безынертной массы»

движения. На входе и на выходе площади сечения определяются из условия минимальных скоростей движения потока. Площадь критического сечения определяется из условия максимальных скоростей движения потока. Максимум скорости ограничивается предельным распределением полной энергии установившегося потока между потенциальной и кинетической по уравнению энергии установившегося вида движения. Для реальных жидкостей и газов эти площади определяются с учётом кинематического коэффициента вязкости.

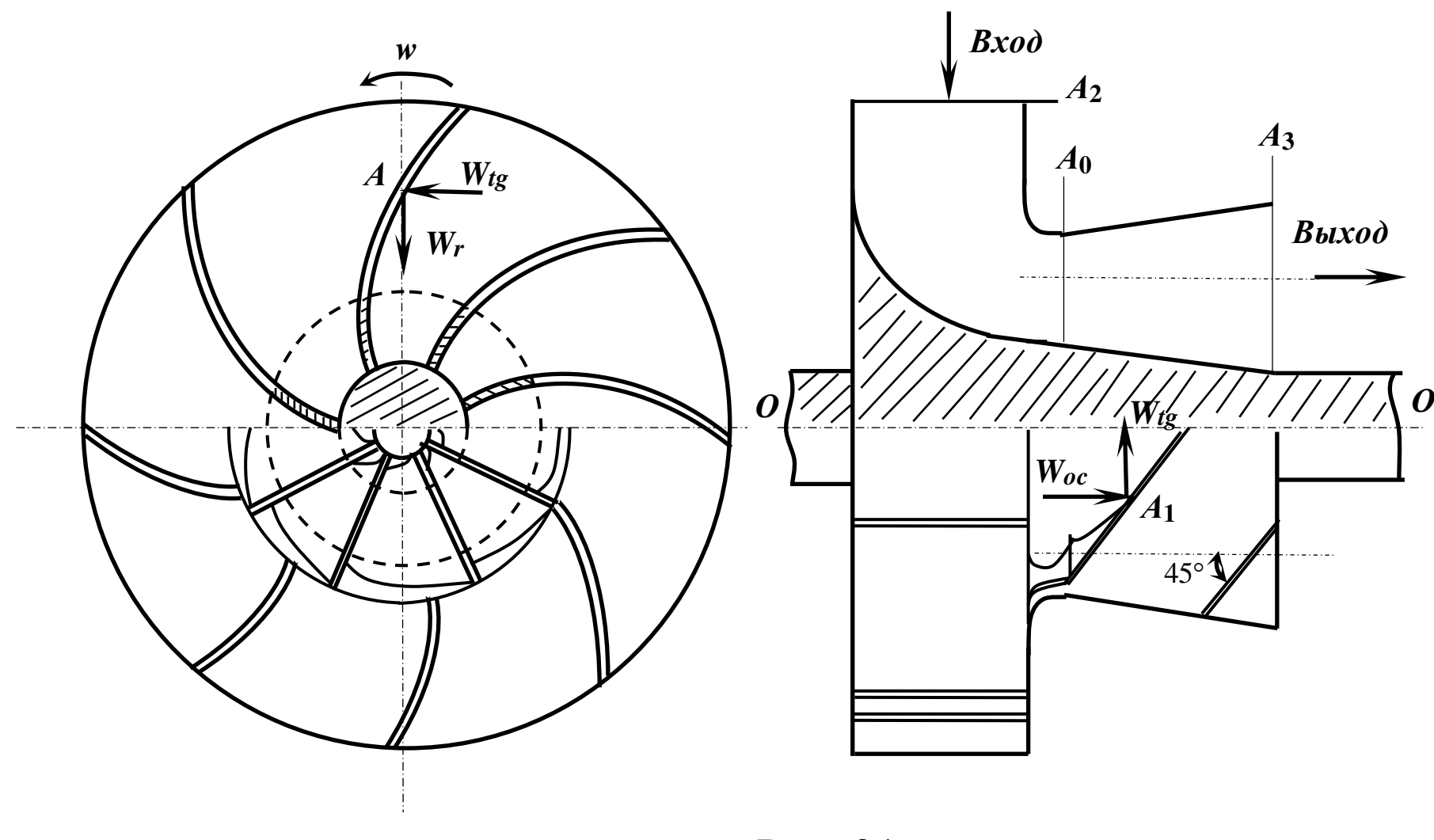

Рис. 31

2.Исходя из полученных сечений, определяем предварительные габариты рабочей полости турбины.

3.Размещаем лопасти турбины в её рабочем объёме. Лопасти турбины размещаются по линиям тока или поверхностям тока.

Для первого участка турбины, который ограничивается входным и критическим сечениями, линиями тока являются логарифмические спирали, лучи которых пересекаются касательными линиями под углом 45°. В радиальном участке лопасти располагаются непосредственно по логарифмической спирали. Затем они переходят в лопасти осевого участка, плоскость которых располагается под углом 45° к осевой скорости потока и перпендикулярно к оси турбины *О-О* (рис. 31). Длина их ограничивается плоскостью критического сечения и плоскостью выхода. Профиль или толщина лопастей турбины должна быть такой, чтобы расчётные площади сечения, входа, выхода и критического сечения, полученные в п.1 данного примера, оставались бы неизменными. Тогда действительные величины площади сечения турбины будут равны расчётным, к которым прибавляются площади сечения лопастей.

Руководствуясь этими сравнительно простыми правилами и положениями, мы получили нужную для наших целей турбину с полным использованием энергии установившегося потока жидкости или газа. Теперь мы должны уяснить себе эти простые правила.

В первую очередь обратим внимание на условия выбора и расположения лопастей турбины. Основными условиями для них являются условия прочности лопастей и условия сохранения рабочей формы.

Условия прочности, с одной стороны, определяются силами механики твёрдого тела, к которым относятся центробежные силы, с другой стороны, – силами механики жидкости и газа. На первом участке такими силами являются тангенциальные силы давления, которые записаны уравнениями (IV-12). Тангенциальные силы давления состоят из статических и динамических сил давления. Тангенциальные статические силы действуют с двух противоположных сторон на лопасти турбины. Поэтому они уравновешивают друг друга и не оказывают силового воздействия на них. Следовательно, только тангенциальные динамические силы давления оказывают силовое воздействие на лопасти турбины. Согласно уравнению (IV-12), величина их на первом участке, от входа в турбину до критического сечения, непрерывно увеличивается. В то же время действительная их величина зависит ещё от числа лопастей турбины. Чтобы выявить их влияние, поступим следующим образом: сначала запишем суммарную, или общую, тангенциальную динамическую силу давления для всего плоского установившегося потока как произведение расхода массы в единицу времени на скорость. Для чего принимаем расход массы как расход массы $M_{tg}$ всего потока в тангенциальном направлении, а скорость как некоторую среднюю величину тангенциальной скорости $W_{tg\text{ ср}}$ для потока первого участка турбины. Тогда получим:

$$R_{tg} = M_{tg} W_{tg\text{ ср}}. \qquad (31)$$

Эта тангенциальная сила $R_{tg}$ не зависти от числа лопастей турбины и является постоянной величиной. Если бы эта величина зависела от числа лопастей турбины, то в зависимости от их числа менялся бы крутящий момент турбины. Чего не происходит в действительности.

Чтобы получить величину тангенциальной силы, которая приходится на долю одной лопасти, мы обязаны разделить эту силу на число лопастей. Обозначим число лопастей через $n$, тогда получим:

$$\frac{R_{tg}}{n} = \frac{M_{tg} W_{tg\text{ ср}}}{n}. \qquad (32)$$

Рассмотрим уравнение (32) по смысловому значению, которое определяет размещение лопастей в объёме турбины. Если мы разделим скорость на число лопастей $n$, то тем самым уменьшим её величину, а это уменьшение приведёт к изменению динамических сил давления для всего потока в целом и непосредственно самого потока. Поэтому делать этого нельзя. Следовательно, на число лопастей можно делить только расход массы в единицу времени. Это значит, что лопасти турбины делят общий поток на ряд потоков, которые в сумме составляют объём расчётного потока без изменения его характеристик.

Уравнения (31) и (32) являются приближенными уравнениями, которыми мы воспользовались для выявления числа лопастей турбины. Поэтому действительную величину тангенциальных сил давления мы должны определять по уравнению (IV-12). Затем полученную величину разделить на число лопастей турбины $n$ и только после этого получим правильную величину тангенциальных динамических сил давления, действующих на лопасти турбины. Тогда уравнение (IV-12) с учётом числа лопастей примет вид:

$$P_{tg\text{ пр.д}} = P_{tg\text{ ст}} + \frac{1}{n}\rho W_{tg}^2. \qquad (33)$$

С точки зрения условий прочности, увеличение числа лопастей турбины приводит к уменьшению величины тангенциальных динамических сил давления. Таким образом, мы определились относительно условий прочности для радиальной ступени турбины.

Отметим, что здесь необходимо обратить внимание на различие действий по отношению к силам между положениями механики твёрдого тела и механики жидкости и газа.

Теперь рассмотрим условие сохранения расчётной формы рабочей полости турбины. В данном примере расчётная форма рабочей полости радиальной турбины соответствует объёму плоского установившегося потока, то есть радиальная полость турбины как бы копирует этот объём. Анализ влияния этих условий делается относительно уравнения энергии установившегося вида движения или уравнения Бернулли. Здесь могут быть два случая.

В первом случае рабочая полость турбины в силу каких-либо причин будет выполнена больше или меньше расчётной. Тогда при меньших размерах полости через турбину пройдёт меньший расход массы, чем это предусмотрено в условиях этого примера, а при больших размерах полости через эту турбину должен пройти больший расход массы, чем это предусмотрено в условиях этого примера. Оба эти случая потребуют изменения полной энергии заданного установившегося потока.

Во втором случае профиль рабочей полости турбины в силу каких-либо причин будет выполнен с отклонением от расчётного. Например, лопасти турбины будут выполнены по логарифмической спирали с большим или меньшим, чем 45°, углом пересечения её лучей с касательными, Это отклонение профиля лопастей приведёт к перераспределению площадей между радиальными и тангенциальными расходами массы в единицу времени, которое потребует либо увеличение, либо уменьшение тангенциальной скорости движения при расчётном общем объёме потока. По этой причине подобные нарушения будут происходить за счёт потерь полной энергии заданного установившегося потока. Величину этих потерь можно определить только экспериментальным путём.

Мы рассмотрели влияние условия сохранения расчётной формы рабочей полости радиальной ступени турбины. Отметим, что расположение лопастей турбины определяет также направление её вращения по часовой стрелке или против неё.

**Рассмотрим теперь осевую ступень турбины**. Для этого нам придётся вспомнить основные положения плоского установившегося вида движения. Согласно положениям плоского установившегося вида движения объём этого потока является цилиндром, в каждой точке которого жидкость движется во взаимно перпендикулярных направлениях. Определяющим расходом потока является радиальный расход массы в единицу времени. Величина скорости этого расхода определяется площадью сечения потока. Площадь сечения потока представляет собой цилиндрическую поверхность. Тангенциальный расход и скорость определяются радиальным расходом массы. Линии тока объединяют всё это движение. Как видим, все его зависимости только в сумме определяют форму плоского установившегося потока, а по раздельности они допускают другие варианты и комбинации своего применения. Выше мы их применяли при разборе установившегося вида движения. Теперь с соответствующими изменениями применим зависимости радиального потока к осевому.

Заменим в осевой ступени турбины радиальный расход массы на осевой $M_{\text{ос}}$. Этот расход является определяющим для осевого потока. Затем, расположив лопасти перпендикулярно к оси потока и под углом 45° к осевой скорости движения $W_{\text{ос}}$, мы тем самым определим направление радиальной скорости движения $W_r$ и радиального расхода массы $M_r$. Для этого потока площади сечения находятся в плоскостях, которые располагаются перпендикулярно оси потока *О-О*. Величина осевой скорости определяется площадью сечения осевого потока. Вот все отличительные особенности осевого потока от радиального. Поэтому все зависимости радиального потока, которые определяют скорости движения и силы давления, пригодны и для осевого потока. В этом случае мы только должны для соответствующих зависимостей заменить знаки, обозначающие радиальное движение, на знаки, обозначающие осевое движение, а тангенциальные зависимости должны остаться без изменения. Если бы, например, в нашем примере радиальная ступень турбины была бы выполнена осевой, то мы все зависимости и положения этой ступени должны были бы оставить без изменений, за исключением линии тока, где мы должны были бы заменить логарифмическую спираль на плоскость тока, которая расположена перпендикулярно оси потока и под углом 45° к осевой скорости движения. Будем считать, что мы рассмотрели первый участок турбины, выполненный в осевом варианте.

Второй участок турбины, где происходит преобразование кинетической энергии в работу, выполнен осевым. Для газов на этой же ступени одновременно происходит преобразовании энергии сжатия в работу. Поэтому увеличение площади сечения потока должно быть выполнено с учётом расширения газа. Только в этом заключается основное различие газовых и жидкостных турбин.

При преобразовании кинетической энергии в работу на лопасти турбины будут оказывать силовое воздействие осевые динамические силы давления. Если бы эта ступень была радиальной, то мы бы назвали эти силы радиальными. Зависимости, как для радиальных, так и для осевых сил давления, одинаковы, поэтому для их записи воспользуемся уравнением (IV-11). Мы его должны переделать с учётом числа лопастей турбины. Тогда действительная величина осевых динамических сил давления, действующих на единицу площади лопасти турбины, будет иметь вид:

$$P_{\text{пр.д}} = P_{\text{ос.ст}} + \frac{1}{n}\rho W_{\text{ос}}^2, \tag{34}$$

где $n$ – число лопастей турбины.

Все остальные положения, которые мы применили для пояснения первого радиального участка, остаются верными и для данного участка турбины. Получив разъяснение для ступеней турбины, рассмотрим её теперь как одно целое.

Силы давления жидкостей и газов действуют в неподвижных плоскостях, поэтому величина их действия не зависит от числа оборотов турбины. Это значит, что величина крутящего момента турбины есть величина постоянная и не зависит от того, вращается ли она или закреплена неподвижно. Отсюда следует, что все необходимые замеры характеристик потока турбины можно делать в неподвижном для неё состоянии, то есть закрепив вал турбины и пропустив через её рабочую полость заданный поток.

Определим коэффициент полезного действия турбины. Он определяется как отношение действительной работы к располагаемой, или теоретической. Запишем его:

$$\eta = \frac{L_д}{L_т}. \tag{35}$$

Теоретическую работу в этом случае можно определить как разность между полной энергией потока и потенциальной энергией среды.

Для жидкости она будет равна:

$$L_т = U - U_{ср}, \tag{36}$$

где $L_т$ – теоретическая работа, $U$ – полная энергия установившегося потока, $U_{ср}$ – потенциальная энергия окружающей среды.

Для газов она будет равна:

$$L_т = U - U_{ср} + \int_1^{V_{ср}} PdV. \tag{37}$$

где $V_{ср}$ – объём газа, расширившегося до состояния среды $V_{ср}$; $\int_1^{V_{ср}} PdV$ – величина энергии сжатия газового потока.

Действительная работа $L_д$ либо замеряется, либо определяется с учётом следующих потерь полной энергии установившегося потока. Здесь учитывается потери, связанные с вязкостью, термодинамическими условиями движения, отклонением рабочей полости турбины от расчётной и так далее. Подставив теоретическую и действительную работу в отношение (35), получим КПД турбины.

Будем считать, что мы пример № 6 выполнили.

Заметим, что полученная турбина, в отличие от современных турбин, не имеет соплового аппарата. Имеет скорости потока на входе и на выходе близкие к нулю. Из этого различия следует, что в современных турбинах совсем не использовали потенциальную и большую часть кинетической энергии установившегося потока, не говоря уже о потерях, связанных с профилированием. Судя по перечисленным недостаткам современных турбин, можно сказать, что их КПД не превышает КПД современных поршневых двигателей. По расчётам данного примера можно практически получить турбину с КПД близким к единице. Отсюда следует вывод, что нет смысла использовать какой-то другой роторный двигатель, который был бы экономичней и проще турбины.

В заключении ещё раз вернёмся к конструктивным особенностям радиальных и осевых турбин. Выше мы рассмотрели их по отношению к зависимостям плоского установившегося вида движения. Теперь рассмотрим по отношению к расположению рабочей полости турбины. Сущность такого расположения заключается в том, чтобы определить, возможно ли, сохранив расчётный объём этих турбин неизменным, разместить его на большем радиусе относительно оси турбины. В этом случае мы сможем сохранить величину динамических сил давления потока неизменными, при этом увеличив радиус их действия относительно оси турбины. Тогда мы сможем получить на турбине больший крутящий момент за счёт увеличения радиуса.

Для радиальной турбины площадью сечения потока является цилиндрическая поверхность, ось которой совпадает с осью турбины. Для осевой турбины плоскость сечения потока лежит в плоскости, которая расположена перпендикулярно к оси турбины и к оси потока. Площадь сечения потока этих турбин определяет величину скорости и величину динамических сил давления, которые непосредственно оказывают силовое воздействие на её лопасти. Поэтому, увеличивая площадь сечения потока, мы тем самым уменьшаем одновременно и скорости, и динамические силы потока. При этом уменьшении динамических сил давления пропорционально квадрату скорости. По этим причинам мы записали условие постоянства плоскостей сечения и объёмов рабочей полости турбины при увеличении радиуса потока.

В рамках этих условий попытаемся увеличить радиус потока. Для радиальной турбины это значит – увеличить одновременно радиус входа, радиус критического сечения и радиус выхода. Для осевой турбины это значит – увеличить радиус цилиндрической поверхности, относительно которого симметрично располагается объём её рабочей полости. Поэтому в радиальной турбине площадь сечения потока находится в прямой зависимости от радиуса потока. Если мы теперь для неё начнём увеличивать радиус потока, то одновременно мы должны будем уменьшать ширину рабочей полости турбины, чтобы сохранить величину площадей сечения неизменными. Для этой турбины нам удастся незначительно увеличить радиус потока, так как ширина полости быстро станет такой, что через неё невозможно будет пропустить необходимый расход реальной вязкой жидкости. Используя эти сравнительно общие рассуждения для радиальной турбины, можно полагать, что в рамках вышеизложенных условий величину радиуса потока можно увеличить незначительно.

Если мы теперь начнём увеличивать радиус потока для осевой турбины, то мы должны будем уменьшать ширину её рабочей полости пропорционально длине цилиндрической поверхности. Исходя из этих общих рассуждений, можно полагать, что величину радиуса потока осевой турбины можно увеличивать на сравнительно большую величину, чем увеличение радиуса радиальной турбины. Следовательно, при определении коэффициента полезного действия необходимо будет учитывать изменение радиуса потока турбин.

Пожалуй, на осевой турбине можно будет получить КПД больше единицы за счёт увеличения радиуса потока. Это положение можно будет выяснить только после более детальных теоретических и экспериментальных исследований.

***Примечание редактора****: к данному примеру редактор приведёт отрывок из текста работы «Строение Солнца и планет солнечной системы с точки зрения механики безынертной массы»:*

Мы упоминали, что коэффициент полезного действия современных конструкций турбин можно увеличить на 30 – 50%. Мы это собирались сделать за счёт того, что газ в турбинах разгоняется до больших скоростей, близких или равных критическим, в неподвижном специальном сопловом аппарате турбины, что позволяет полностью использовать потенциальную часть энергии потока.<…>

Другой большой статьей потерь, которую можно устранить, является проточная полость турбины, геометрические размеры которой сравнительно далеки от необходимых естественных форм, предписанных законами природы. <…> Если ещё учесть, что в большинстве своём турбины применяются как двигатели в совокупности с насосами и компрессорами, на привод которых тоже затрачивается энергия турбины, то коэффициент полезного действия такого сотрудничества ещё возрастет за счёт увеличения коэффициента полезного действия насоса или компрессора. Следовательно, турбокомпрессионные двигатели и просто турбины, изготовленные в соответствии с естественными законами движения жидкостей и газов, представляют собой идеальный двигатель с максимально возможным коэффициентом полезного действия, который только можно получить практическим путем. Подобный двигатель может с большим успехом заменить даже поршневые двигатели любых мощностей и систем.<…> Газотурбинный двигатель, построенный по законам механики безынертной массы, поможет устранить курьез сегодняшнего дня. Всем известен современный газотурбинный двигатель, рабочим ходом которого является вращательное движение, а изобретатели всего мира ищут двигатель с вращательным рабочим ходом, который бы происходил по принципу поршневого двигателя. Например, таким двигателем является двигатель Ванкеля. Вы спросите, не чудаки ли они? Нет, конечно. Современные турбины с их вышеперечисленными недостатками невозможно применить в тех областях техники, где сейчас применяются поршневые двигатели. По этой причине поршневые двигатели с таким большим недостатком, который состоит в том, что рабочее возвратно-поступательное движение их поршней преобразуется во вращательное движение с помощью довольно быстроизнашивающегося кривошипо-шатунного механизма, остаются вне конкуренции для турбин в своей области.<…> Остаётся добавить, что газотурбинные двигатели нового типа будут более гигиеничными из-за своего высокого к.п.д. Да и шум, производимый двигателями, тоже устраняется вместе с приводной трансмиссией типа коробок передач, карданных валов и т.д. В данном случае всю эту трансмиссию можно заменить гидравлической трансмиссией. (С)

## ПРИМЕР № 7

Требуется определить истечение жидкости и газа из круглого отверстия от действия статических сил давления.

Здесь имеются в виду отверстия, которые расположены в стенках объёмов, в которых жидкости и газы находятся в состоянии покоя под действием определённых силовых полей, или отверстия в стенках трубопроводов, где движется определённый установившийся поток. Покажем на рисунке 32 такое отверстие, с вытекающей из него жидкостью или газом. При вытекании жидкости или газа из отверстия образуются как бы два потока от действия разности сил давления окружающей среды и определённого объёма жидкости или газа.

Непосредственно у стенки объёма образуется поток плоского установившегося вида движения (рис. 32). В этом потоке жидкости и газы преобразуют потенциальную энергию в кинетическую. На внутреннюю границу плоского установившегося потока, который совпадает с диаметром отверстия в стенке (рис. 32) жидкости и газы поступают, имея только кинетическую энергию. С внутренней границы потока к оси отверстия они движутся по линиям тока, которые имеют форму окружности с определённым радиусом. По такой форме линий тока нет потерь энергии. Таким образом, происходит формирование границ струи жидкости или газа с внутренней стороны стенки относительно оси отверстия. Затем жидкость или газ поступает в отверстие стенки уже в виде сформировавшегося потока и дальше – во внешнюю среду.

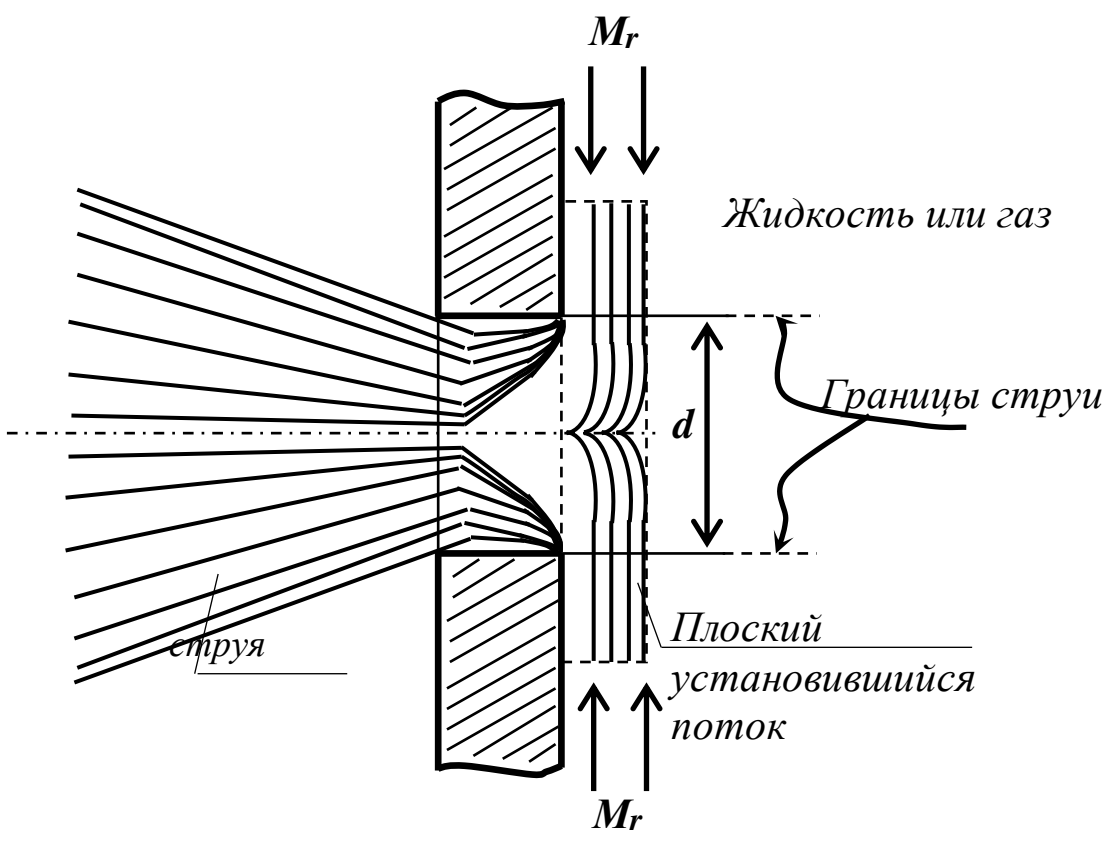

Рис. 32

Возьмём в качестве плоскости исследования струи внутреннюю поверхность стенки и запишем необходимые зависимости. Тогда уравнением движения струи будет уравнение движения установившегося вида движения. За площадь сечения струи принимаем площадь отверстия с диаметром $d$. Запишем его:

$$M_{\text{с}} = \rho W \frac{\pi d^2}{4}. \tag{38}$$

В этом сечении жидкости и газы не обладают статическим давлением, так как не имеют потенциальной энергии. По этой причине мы должны будем записать для струи только уравнение динамических сил давления. Оно будет иметь вид:

$$P_{\text{дин}} = \frac{4M_{\text{с}}}{\pi d^2} \cdot W. \tag{39}$$

Воспользовавшись уравнениями (38) и (39), мы найдем все необходимые характеристики струи. Характеристики плоского установившегося потока определяются по зависимостям плоского установившегося вида движения. Поэтому мы их здесь не будем переписывать, так как каждый для себя сможет записать их самостоятельно.

В современной литературе в отношении струи предпочитают иметь зависимость расхода массы непосредственно от статических сил давления. Мы тоже постараемся получить подобную зависимость. Для чего заменим динамические силы давления $P_{дин}$ в уравнении (39) через разность статических сил давления среды $P_{ст.ср}$ и объёма $P_{ст}$. Затем заменим скорость $W$ в уравнении (38) по уравнению (39), тогда получим:

$$M_{с} = \frac{\pi d^2}{4}\sqrt{\rho\left(P_{ст} - P_{ст.ср}\right)}. \qquad (40)$$

Подобное уравнение называют уравнением расхода отверстия.

При определении расхода массы для реальных вязких жидкостей и газов мы обязаны умножить правую часть уравнения (40) на коэффициент кинематической вязкости μ, тогда оно примет вид:

$$M_{с} = \mu\frac{\pi d^2}{4}\sqrt{\rho\left(P_{ст} - P_{ст.ср}\right)}. \qquad (41)$$

Уравнение (41) является уравнением расхода реальных вязких жидкостей и газов для отверстия круглой формы.

На практике часто применяют круглые отверстия с различной формой внутренней стороны стенки. Подобные отверстия могут иметь расход массы отличный от расхода массы отверстия с такой же площадью в плоской стенке. Подобное различие можно учесть с помощью коэффициента формы отверстия ξ. Запишем его как отношение расхода массы $M_{ф}$ отверстия с иной формой к расходу массы отверстия в плоской стенке $M_{с}$, получим:

$$\xi = \frac{M_{ф}}{M_{с}}. \qquad (42)$$

Этот коэффициент проще всего определить экспериментальным путем и свести его величины для характерных отверстий в таблицы. Тогда более общее уравнение расхода будет иметь вид:

$$M_{с} = \xi\mu\frac{\pi d^2}{4}\sqrt{\rho\left(P_{ст} - P_{ст.ср}\right)}. \qquad (43)$$

Уравнение (43) также пригодно для отверстий иных форм, например, квадратных, треугольных и так далее. Для них тоже необходимо будет ввести свой коэффициент формы отверстия.

Заметим, что в современной практике применяется уравнение расхода такого вида:

$$M_{с} = \mu\frac{\pi d^2}{4}\sqrt{2\rho\left(P_{ст} - P_{ст.ср}\right)}.$$

где μ называется коэффициентом расхода.

Уравнение в таком виде записано для невязкой жидкости. Поэтому сравним его с уравнением (40), которое тоже записано для невязкой жидкости. Оно отличается от уравнения (40) тем, что имеет коэффициент расхода μ и цифру 2 под корнем. Такое различие вызвано тем, что при выводе этого уравнения пользовались уравнением Бернулли, а не уравнением сил. Поэтому под корнем появилась двойка. Чтобы компенсировать эту ошибку, ввели в это же уравнение коэффициент расхода μ, а его появление объяснили так называемым «сужением» струи.

Мы рассмотрели уже семь примеров, которые охватили состояние покоя жидкостей и газов, установившегося и плоского установившегося вида движения. Согласно порядку изложения следующий пример должен быть по расходному виду движения жидкостей и газов. В данной работе мы не будем его приводить, а сошлемся на пример № 3 работы [3], который является наиболее характерным для данного вида движения[10]. Дальше перейдем к акустическому виду движения жидкости и газа.

---

[10] автор даёт ссылку на заявку от 21.12.1969 г. на предполагаемое открытие «Расходное движение жидкости и газа». В настоящее время редактор не имеет возможности опубликовать текст заявки, т.к. архив в этой части не подготовлен к публикации.

## ПРИМЕР № 8

Требуется рассчитать глушитель для глушения выхлопа газов на выходе из поршневых двигателей.

Непосредственно от конкретного двигателя для этой цели мы должны будем получить следующие данные:

массу газа на один выхлоп $m$, температуру $T_в$, давление $P_в$, плотность $\rho_в$ при начале выхлопа, величину времени выхлопа $t_в$, площадь выхлопного окна $F_в$.

Затем берут зависимости акустического вида движения и приступают к расчёту необходимого глушителя. Всё очень просто, но не совсем. Прежде, чем приступить к расчёту, необходимо знать ещё хотя бы конструктивную схему глушителя и принцип его работы. Эту схему и принцип работы можно получить непосредственно из знания природы и характера акустического вида движения жидкостей и газов. Для этого недостаточно знать зависимости этого вида движения. Необходимо сравнительно хорошо представлять себе картину движения при акустическом виде движения. Поэтому сначала постараемся уяснить её себе. При этом нам придётся вспомнить свойства реальных жидкостей и газов.

Затем берём пластину источника возмущения конечной определённой площади $F$. Дальше полагаем, что эта пластина совершила поступательное движение из положения I в положение II сразу на всю длину своего хода $l$, а затем прекратила своё движение. Покажем это на рис. 33.

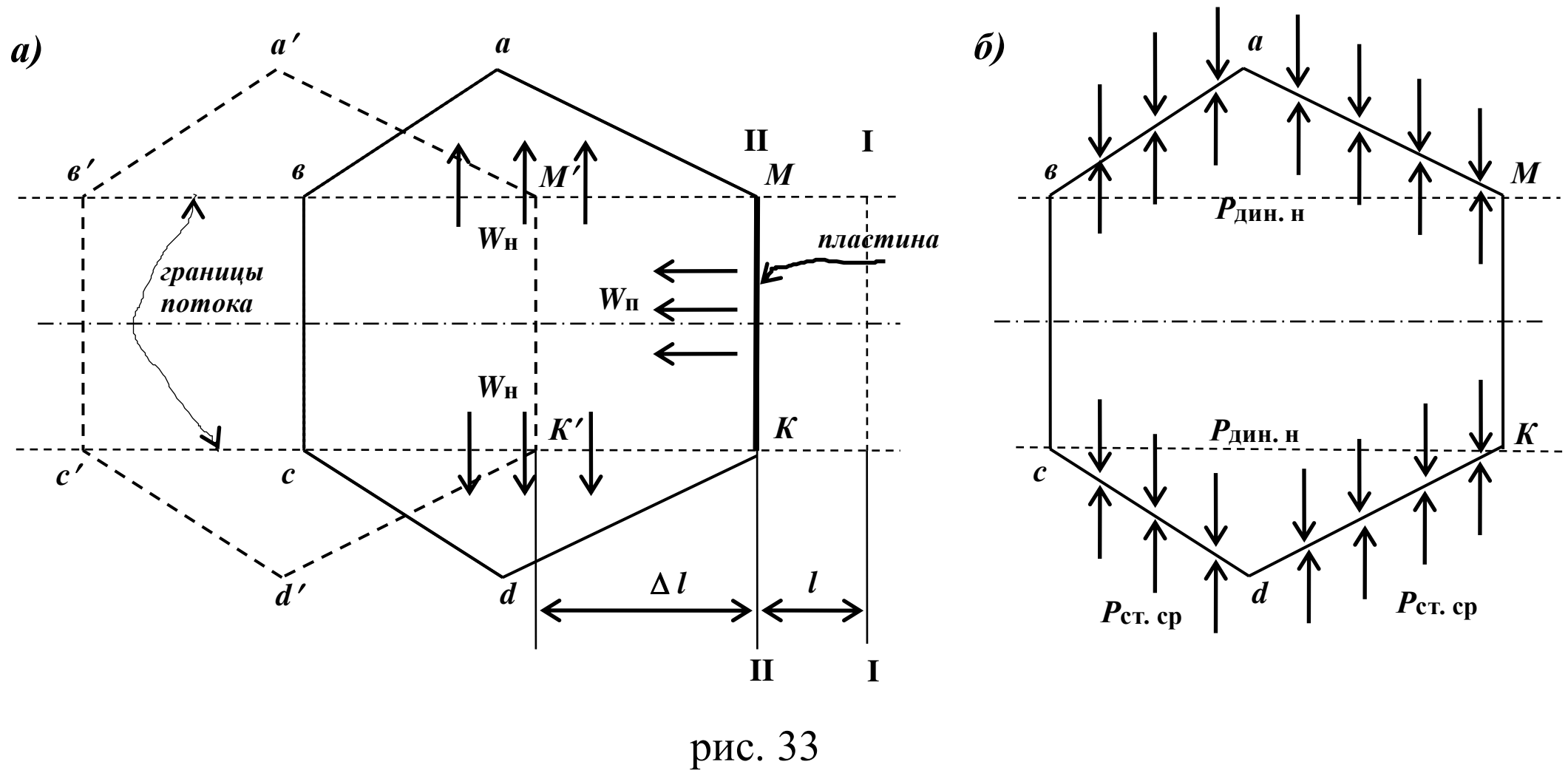

рис. 33

При своём перемещении пластина вытеснит определённый объём $V$, который поступает в окружающую среду. Считаем время останова пластины в положении II фиксированным.

К этому времени вытесненный газа займет в окружающей среде некоторый объём «*а, в, с, d, К, М*», который мы назвали зоной возмущения. Он представляет собой объём тела вращения, осью которого служит ось потока. Поверхности объёма: «*а, в, М* и *с, d, К*» могут быть любой другой формы, чем той, которая показана на рис. 33. Форма определяется скоростью движения пластины возмущения. Здесь они показаны треугольной формы лишь для удобства изображения. Поступательные скорости движения и силы давления в этом объёме определяются зависимостями акустического вида движения.

Мы знаем, что образовавшийся объём «*а, в, с, d, К, М*» с течением времени как-то должен изменяться. Чтобы определить это изменение, полагаем, что изменение происходит в течение небольшого промежутка времени $\Delta t$ от первоначального фиксированного времени. Через время $\Delta t$ берём второй фиксированный момент движения и смотрим, что произошло к этому времени.

К этому времени объём «*а, в, с, d, К, М*» просто переместится на некоторое расстояние $\Delta l$ по прямой в направлении акустического движения и займет новое положение «*а′, в′, с′, d′, К′, М′*». При этом величина и форма объёма останутся неизменными. Если мы рассмотрим ещё целый ряд фиксированных моментов, начиная от первого, то мы заметим, что возмущённый объём просто перемещается со скоростью возмущения или скоростью звука. При этом до и после движущегося объёма остаётся зона невозмущённой среды. Теперь мы можем объёмно представить себе акустический поток жидкости или газа.

Дальше мы должны уяснить, как образуется и каким способом сохраняется движущийся объём возмущённой жидкости или газа. Поступательный расход массы в единицу времени, который определяется движением пластины возмущения, действует в объёме потока. Границы потока показаны на рис. 33, *а*. Этот расход вытесняет из объёма потока нормальный расход массы, который движется перпендикулярно границам потока. Поступательный и нормальный расходы массы в единицу времени равны между собой. Определяющим среди них является поступательный расход. Нормальные скорости зависят не только от поступательного расхода, но и от площади сечения. Для нормального потока она является переменной. За время движения пластины из положения I в положение II высота площади сечения потока увеличивается пропорционально скорости возмущения или скорости движения звука. И поступательное, и нормальное движение определяется зависимостями акустического вида движения. Мы просто вспомнили связь между поступательным и нормальным движением акустического вида движения.

Под действием нормальных динамических сил давления происходит изменение границ потока, например, до нового положения: *а, в, М* (рис. 33, *а*), так как они действуют нормально к границам потока. Новое положение границ *а, в, М* определяется тем, что в начальный момент на границе потока нормальные динамические силы давления бывают больше, чем статические силы окружающей среды. Поэтому границы зоны возмущения начинают увеличиваться. Но с ростом площади границ происходит уменьшение нормальных динамических сил давления. Это уменьшение находится в прямой зависимости от увеличения площади. Затем наступает такой момент, когда нормальные динамические силы давления станут равными статическим силам давления окружающей среды. В нашем примере такой момент наступает на границах объёма *а, в, М* и *с, d, К*. Таким образом определяются границы объёма возмущения в нормальном движению потока направлении, а изменение его границ в направлении движения нам уже известно. Зависимостей для определения нормальных границ возмущения в данной работе нет, но их легко можно установить из условия равновесия статических сил давления и нормальных динамических сил давления. Следовательно, величина возмущённого объёма поддерживается динамическими силами давления, которые уравновешиваются статическими силами давления окружающей среды.

Динамические силы давления сохраняются в объёме за счёт перемещения всего возмущённого объёмаа со скоростью возмущения, так как для их существования необходим соответствующий расход массы в единицу времени. В конечном итоге, состояние возмущённого объёма определяется избытком энергии по сравнению с окружающей средой, который получается за счёт работы пластины возмущения. Величина избытка энергии определяется уравнением работ. По этому уравнению мы можем получить её количественную величину. Дальше для реальных жидкостей и газов мы обязаны записать избыток энергии уравнением термодинамической энергии сжатия VII-7, так как он приводит к изменению термодинамического состояния среды. По уравнению работ V-46 мы вычислим величину энергии сжатия. Тогда по уравнению VII-7 останется определить пределы изменения объёма газа в возмущённой среде, то есть пределы интеграла этого уравнения. Для этого нам ещё потребуется зависимость адиабатического процесса, так как избыток энергии в окружающей среде мы получаем за счёт подвода в неё определённого количества массы в виде поступательного расхода массы в единицу времени. В этом примере мы подчеркнули другую особенность содержания избытка энергии в среде, которая отличается от избыточного содержания энергии набегающего потока.

Отметим, что для образования волны достаточно одного этапа движения пластины, например, из положения I в положение II, хотя мы выше определили длину волны как образованную за два этапа движения пластины возмущения. Это было сделано для того, чтобы подчеркнуть различие в длине волн первого и второго этапов движения, а не в угоду ныне существующему понятию о длине волны, которое, пожалуй, было принято лишь в угоду синусу, а не самой природе акустического движения.

Если бы мы теперь начали рассматривать движение пластины из положения II в положение I, то нам бы пришлось повторить все вышеизложенные рассуждения, с той лишь разницей, что они должны были бы относиться к возмущённому объёму с недостатком энергии относительно окружающей среды. В связи с этим скорости изменят свой знак на противоположный относительно скоростей возмущённого объёма первого этапа движения.

Мы несколько отвлеклись от задачи нашего примера. Мы просто должны найти нужную нам конструктивную схему глушителя и принцип его работы, так как движение газа при выхлопе соответствует акустическому движению, которое определяется как движение пластины конечной площади из положения I в положение II.

Дополнительная величина энергии возмущённого объёма относительно энергии окружающей среды создаётся путем ввода в этот объём дополнительной массы. Если убрать этот излишек массы, полностью или частично, то дополнительная энергия возмущённого объёма тоже либо исчезнет, либо останется частично.

Возьмём это положение в качестве принципа работы глушителя выхлопа газов. Следовательно, конструктивная схема глушителя должна быть такой, чтобы она могла обеспечить необходимый расход массы из возмущённого объёма. В конечном итоге необходимо убрать дополнительную энергию возмущённого объёма. Величина механической энергии равна произведению объёма на давление. Тогда количество изъятой энергии мы можем определить как произведение объёма глушителя $V_{гл}$ на величину энергии единицы объёма возмущённой части среды. Величина энергии единицы объёма возмущённой зоны определяется уравнением V-47.

Тогда изъятое количество энергии $U_{из}$ можно записать в такой форме:

$$U_{из} = V_{гл}\left(U + \nu\rho \int_{W_{п1}}^{0} W_{п}\, dW\right) \qquad (44)$$

Величина изъятой энергии должна быть равна или быть меньше дополнительной величины энергии $U_{доп}$ возмущённого объёма. Количество дополнительной энергии $U_{доп}$ определяется как произведение объёма газа $V_{газ}$ на работу единицы объёма. Эта работа определяется уравнением V-46. Тогда количество дополнительной энергии $U_{доп}$ можно записать в таком виде:

$$U_{доп} = V_{газ} \cdot \nu\rho \int_{W_{п1}}^{0} W_{п}\, dW\,. \qquad (45)$$

Теперь, приняв энергию $U_{из}$, изъятую объёмом глушителя, равной или меньшей дополнительной энергии $U_{доп}$ , мы сможем по уравнению (44) определить объём глушителя $V_{гл}$, необходимый для изъятия определённого количества дополнительной энергии.

Мы получили нужный объём глушителя. Теперь мы должны оформить его конструктивно таким образом, чтобы он соответствовал принципу работы глушителя.

Покажем на рис. 34 одну из схем конструктивного оформления глушителя. Непосредственно выхлопным патрубком глушитель соединен с выхлопным окном цилиндра поршневого двигателя.

Объём глушителя образует прямоугольник, вытянутый вдоль оси. В стенках глушителя просверлено множество отверстий, которые прикрываются лепестковыми клапанами. Задача этих лепестковых клапанов заключается в том, чтобы выпускать газ из объёма глушителя и не пропускать газ в его . На рис. 34 показано, что лепестковый клапан имеет ограничитель хода. Это несущественное дополнение. Рассмотрим работу глушителя.

При выхлопе газы попадают в объём глушителя через выхлопной патрубок. Через площадь сечения выхлопного патрубка плоскости $S_1$ выхлопные газы будут поступать в объём глушителя как поступательный расход массы в единицу времени для акустического вида движения при выталкивании некоторого объёма пластиной возмущения. Этот расход начнёт выталкивать в нормальном направлении соответствующий расход массы газа, который к этому моменту заполнял объём глушителя.

Через отверстия в стенках глушителя нормальный расход начнёт поступать в окружающую среду. Клапана не препятствуют этому расходу. В результате образуется объём возмущённой среды, частью которого является объём глушителя. Возмущённый объём начнёт перемещаться в среде в соответствии с направлением акустического движения. Часть возмущённого объёма, которая заключена в объёме глушителя, останется без движения, так как лепестковые клапана закрываются и прерывают связь этих объёмов. Поэтому избыточная часть энергии остаётся в объёме глушителя. Без дополнительной энергии возмущённый объем перестает быть возмущённым объёмом. Следовательно, при выхлопе в зоне глушителя образуется акустический объём возмущённой среды, но он не перемещается в пространстве, так как глушитель лишает его дополнительной энергии. Как видим, конструктивная схема глушителя (рис. 34) выполняет функции своей работы.

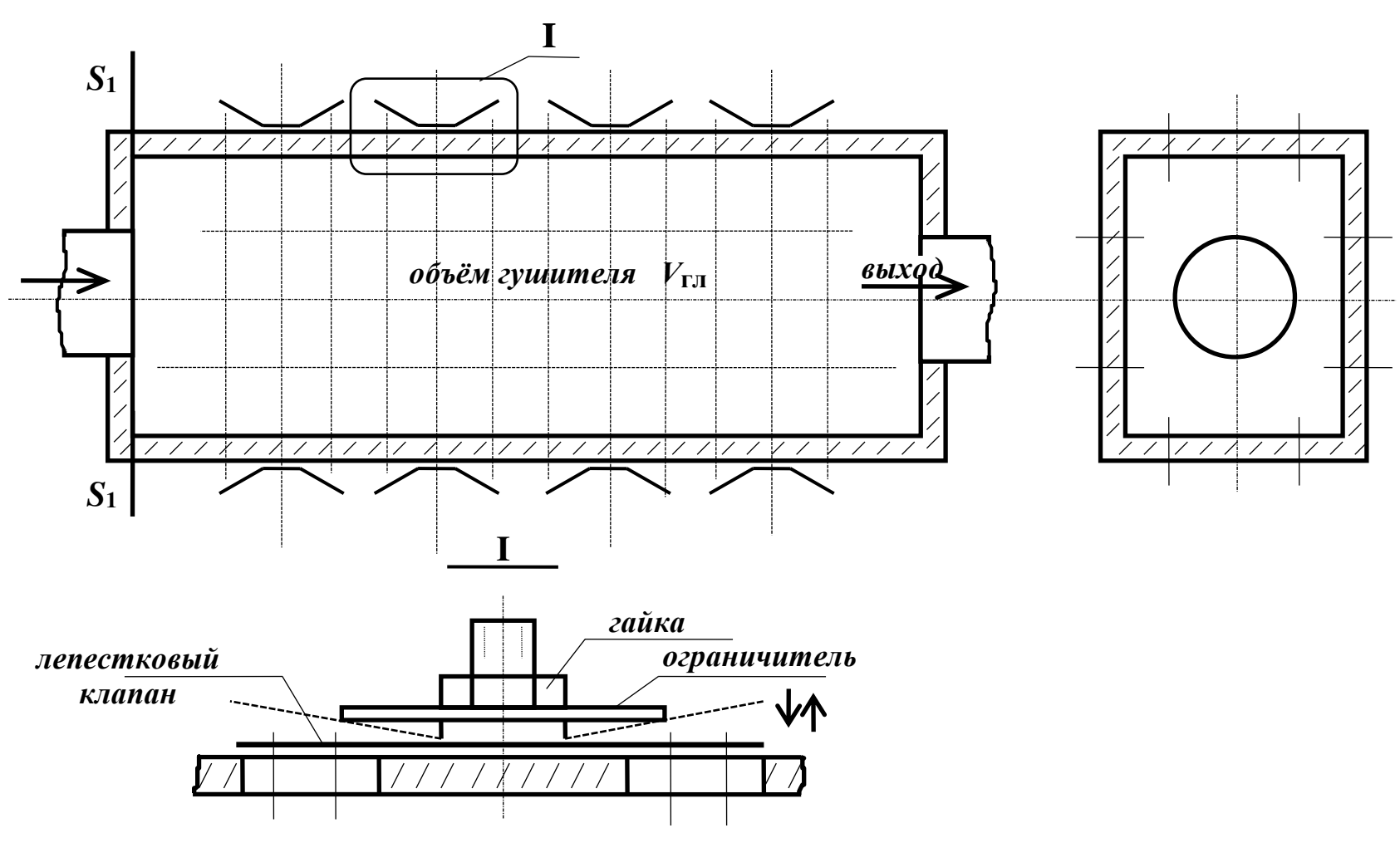

Рис. 34

Остаётся определить остальные характеристики глушителя. Стенки глушителя должны располагаться как можно ближе к площади сечения входного патрубка или выхлопного окна цилиндра. После этого определяется длина глушителя.

Общая площадь отверстий в стенках глушителя определяется расходом массы выхлопных газов. Этот расход попадает в объём глушителя и выталкивает нормальный расход массы, который по величине равен расходу массы выхлопных газов. По уравнениям акустического вида движения вычисляются нормальные скорости и нормальные динамические силы давления. По величине нормального расхода и нормальной скорости определяются динамические силы давления. По величине нормального расхода и нормальной скорости с помощью уравнений движения определяется площадь всех отверстий в стенках глушителя.

Нормальные динамические силы давления открывают лепестковые клапана. Поэтому силу упругости лепестковых клапанов надо рассчитывать на их действие. Проделав эти несложные расчёты, мы определим необходимые характеристики глушителя.[11]

Отметим, что работа глушителя подобного типа не уменьшает мощность поршневых двигателей, как это делают глушители современных конструкций.

***

## Расчёт крыльчатки центробежного насоса

*(из рукописи «Строение Солнца и планет солнечной системы с точки зрения механики безынертной массы» (1973 – 1974 г.г.)*

Чтобы окончательно уяснить себе положения механики безынертной массы, разберём один практический пример, который пригодится нам в будущем. Этим примером будет расчёт центробежного насоса.

Центробежные насосы, как и любые другие насосы, предназначены для перекачки жидкости. Устройство их очень простое. Они состоят из вращающегося колеса или, как его ещё называют, крыльчатки и неподвижного корпуса, который имеет форму улитки. Лопатки на колесе закреплены между двумя дисками. Ось колеса соединена с электродвигателем, который приводит колесо во вращательное движение. Жидкость начинает поступать на вход насоса. Затем она лопатками колеса перебрасывается в корпус насоса и выходит из него с новыми энергетическими параметрами. Вот, собственно, и всё устройство насоса и его рабочее движение.

Для проектирования насоса задают следующие параметры:

1. жидкость, которую необходимо перекачивать. Следовательно, известна плотность $\rho$ этой жидкости;

[11] насколько редактор помнит, автор сказал, что «цунами – это перевёрнутая волна». Редактор считает, что будет возможно создать «глушители» и для такой формы акустического движения, как цунами, для защиты густонаселённых побережий.

2. энергетические характеристики потока жидкости на входе в насос в виде полного давления $P_{\text{пол.вх}}$.

В этом случае поток жидкости на входе надо понимать как установившийся вид движения жидкости. Поэтому для него надо применять зависимости установившегося вида движения. Тогда заданное полное давление потока $P_{\text{пол.вх}}$ надо понимать как количество энергии единицы объёма входного потока:

$$E_{\text{ед}} = P_{\text{пол.вх}}\,(1/м^3).$$

Такой вид будет иметь энергетический уровень входного потока.

3. Количество перекачиваемой жидкости в виде объёма, перекачиваемого насосом в единицу времени. Обозначим этот объём, как объём потока перекачиваемой жидкости $V_{\text{пот}}$. Обычно задают количество перекачиваемой жидкости за час работы насоса, но мы будем считать, что нам задали объёмный расход жидкости за одну секунду работы насоса.

4. Задают также энергетический уровень потока на выходе из насоса в виде полного давления потока на выходе из насоса $P_{\text{пол.вых}}$. Его надо понимать так же, то есть как количество энергии единицы объёма жидкости на выходе из насоса.

Мы получили все необходимые параметры для проектирования центробежного насоса. Нам остаётся выбрать электродвигатель, соответствующий по своей мощности производительности насоса, и получить все необходимые геометрические размеры колеса, или крыльчатки насоса, по которым можно было бы изготовить её для насоса заданной производительности.

Электродвигатели изготавливаются нашей промышленностью с различным числом оборотов $n$ и различной мощности $N$. Это различие моторов принято в пределах определённых интервалов. Поэтому они занесены в каталог. Получив необходимую мощность насоса и число оборотов для него, мы сможем выбрать по этим характеристикам соответствующий тип электродвигателя.

Приступаем к необходимым расчётам для проектирования насоса.

***Выбор электродвигателя насоса***

Для этого мы должны проделать следующие действия.

Определяем мощность насоса по энергетическому уровню потоков.

Работа и энергия определяются как произведение объёма на давление. Объём потока на выходе из насоса нам задан, на входе он будет таким же, т. к. плотность жидкости является постоянной величиной. Полное давление потока на входе и на выходе из насоса нам тоже задано. Насос должен обеспечить прирост уровня энергии входного потока до уровня выходного потока. Тогда потребляемая насосом энергия или работа запишется в таком виде:

$$E_{\text{нас}} = VP_{\text{пол.вых}} - VP_{\text{пол.вх,}} \quad \text{или}$$

$$E_{\text{нас}} = V(P_{\text{пол.вых}} - P_{\text{пол.вх}}). \tag{1}$$

Но у нас объём потока задан за единицу времени, которая равна одной секунде. Это значит, что в уравнении (1) мы должны заменить объём потока $V$ на секундный объёмный расход жидкости $V_{\text{пот}}$. Тогда уравнение (1) даст нам работу насоса в единицу времени, то есть определит нам мощность. Поэтому уравнение (1) после такого преобразования будет выражать мощность насоса:

$$N_{\text{нас}} = V_{\text{пот}}(P_{\text{пол.вых}} - P_{\text{пол.вх}}) \tag{2}$$

Уравнение (2) выражает потребляемую насосом мощность. Определив величину мощности, мы подберём необходимый для нашего насоса электродвигатель. Из каталога электродвигателей выбираем электродвигатель нужной мощности с возможно большим числом оборотов $n$. После выбора по каталогу электродвигателя помимо мощности нам будет известно число оборотов его вала в единицу времени. На этом предварительный выбор электродвигателя заканчивается.

***Нахождение нужных геометрических размеров крыльчатки***

Для этого нам прежде необходимо определиться с видом движения жидкости в крыльчатке насоса. Этим видом движения будет плоский установившийся вид движения жидкости. Поэтому для поисков необходимых геометрических размеров нам придётся воспользоваться зависимостями именно этого вида движения. Далее мы будем использовать эти зависимости в такой последовательности, чтобы, исходя из уже известных величин, постепенно выявить искомые величины.

Прежде, чем приступить к поискам необходимых величин, отметим особенности плоского установившегося потока жидкости в полости центробежного насоса. Одна его особенность состоит в том, что он должен увеличить энергетический уровень потока с меньшим энергетическим уровнем. Вторая особенность заключается в том, что ввод энергии осуществляется лопатками колеса насоса при их движении в тангенциальной плоскости сечения потока. Кроме того, непосредственно сам объём плоского установившегося потока жидкости располагается по своим границам в пределах границ колеса крыльчатки насоса, то есть геометрические размеры крыльчатки определяют геометрические размеры потока. Следствием этих особенностей является то положение, что поверхности лопастей колеса насоса должны совпадать с поверхностями тока плоского установившегося движения жидкости. Поэтому, исходя из этого положения, мы должны, в первую очередь, определить необходимые характеристики потока в тангенциальном направлении движения и в последующем исходить из них.

***1. Найдем тангенциальную площадь сечения потока***

Прирост энергии потока в этом случае происходит за счёт работы вытеснения жидкости в тангенциальном направлении лопатками колеса. Поэтому, как и в прямолинейном потоке первого этапа движения поршня при акустическом виде движения жидкости работа поршня переходит в энергию прямолинейного потока, так и в тангенциальном потоке плоского установившегося движения жидкости работа лопаток колеса переходит в энергию тангенциального потока. Тем самым создаётся рост энергии потока.

В радиальном направлении движение жидкости создаётся путем перераспределения потенциальной и кинетической энергии радиального потока, полная энергия которого в этом случае будет равна тому приросту энергии, который получается в тангенциальном потоке. Движение жидкости в радиальном направлении происходит способом, аналогичным движению жидкости в нормальном направлении акустического потока жидкости на первом этапе движения поршня. По этой причине скорости движения потока в тангенциальном направлении должны быть равны скорости движения лопаток колеса насоса, а скорости движения потока в радиальном направлении будут зависеть от площадей сечения потока в радиальном направлении. Для тангенциальной площади сечения потока, как мы знаем, величина её является постоянной, а тангенциальные скорости движения потока по всей площади его сечения одинаковы и равны по величине. Величина прироста полного давления нам известна как разность полных давлений на выходе и входе в насос, то есть

$$P_{\text{прир}} = P_{\text{пол.вых}} - P_{\text{пол.вх}}. \tag{3}$$

Величины этих полных давлений нам заданы. Как и при акустическом движении жидкости, величина прироста полного давления потока будет равна величине динамического давления, создаваемого скоростью движения лопаток колеса. Динамическое давление лопаток колеса будет равно:

$$P_{\text{кол}\,tg} = \rho W^2_{\text{кол}\,tg}. \tag{4}$$

В уравнении (4) динамическое давление потока $P_{\text{кол}.tg}$ нам известно по уравнению (3). Плотность жидкости нам тоже известна заданным условием. Тангенциальная скорость движения лопастей крыльчатки нам неизвестна. Поэтому подставим в уравнение (4) величину прироста полного давления из уравнения (3) и найдем искомую скорость:

$$W_{\text{кол}\,tg} = \sqrt{\frac{P_{\text{пол.вых}} - P_{\text{пол.вх}}}{\rho}}. \tag{5}$$

По уравнению (5) мы определяем тангенциальную скорость колеса, которая должна быть равна тангенциальной скорости потока, то есть

$$W_{\text{кол}\,tg} = W_{\text{пот}\,tg}. \tag{6}$$

Зная тангенциальную скорость потока $W_{пот.tg}$, мы можем из тангенциального уравнения движения потока, которое имеет вид:

$$M_{tg} = \rho F_{tg} W_{пот\ tg}, \qquad (7)$$

найти тангенциальную площадь сечения потока, т.к. расход массы в единицу времени $M_{tg}$ нам фактически задан в условиях проектирования. Он будет равен произведению объёмного заданного расхода жидкости $V_{пот}$ на её плотность, то есть

$$M_{tg} = \rho V_{пот}. \qquad (8)$$

Подставив значение расхода массы (8) в уравнение (7), мы определим величину тангенциальной площади сечения потока. Она будет равна:

$$F_{tg} = \frac{V_{пот}}{W_{пот\,tg}}. \qquad (9)$$

Мы определили все необходимые характеристики, связанные с тангенциальным потоком. Остаётся распределить их для потока.

Возможно, у вас возник вопрос: «Почему, если лопатки колеса по своей длине от центра потока к его периферийным границам имеют разные линейные скорости, мы приняли тангенциальную скорость потока постоянной?» Это законный вопрос. Постоянство тангенциальных скоростей определяется кривой линией потока. Если вы хорошо подумаете, то теперь сможете сами ответить на этот вопрос более подробно. И не задавайте подобных вопросов лицам, занимающимся механикой твёрдого тела [12]

***2. Найдем геометрические размеры крыльчатки насоса***

Для чего сначала найденную площадь сечения $F_{tg}$ представим в виде прямоугольника с длиной $l$ и шириной $h$ (рис. 1), то есть

$$F_{tg} = lh. \qquad (10)$$

Длина и ширина тангенциальной площади сечения в первом приближении выбираются просто из конструктивных соображений. Затем мы постараемся найти окружность, где величина тангенциальных скоростей $W_{tg}$ потока совпадает с линейной скоростью вращения окружности с соответствующим радиусом. Линейная скорость $W_{окр}$ будет равна длине окружности ($2\pi r$), умноженной на число оборотов вала электродвигателя $n$, которое мы получили при выборе электродвигателя, то есть

$$W_{окр} = 2\pi r n. \qquad (11)$$

Затем подставляем в уравнение (11) вместо окружной скорости $W_{окр}$ величину тангенциальной скорости потока $W_{кол\,tg}$, найденную по уравнению (5), получим:

$$W_{кол\,tg} = \sqrt{\frac{P_{пол.вых} - P_{пол.вх}}{\rho}} = 2\pi r n. \qquad (12)$$

В уравнении (12) нам известны все величины, кроме радиуса окружности $r$. Поэтому мы решим это уравнение относительно радиуса $r$, получим:

$$r = \frac{W_{кол\,tg}}{2\pi n}. \qquad (13)$$

[12] расчёт насоса взят редактором из работы «Строение Солнца и планет солнечной системы с точки зрения механики безынертной массы», поэтому местоимение «вы», в данном случае, является обращением автора к своим детям, которым адресован данный труд.

На окружности, с полученным по уравнению (13) радиусом $r$, должны совпадать не только тангенциальные $W_{\text{кол}\ tg}$ и окружные $W_{\text{окр}}$ скорости потока, но и радиальные скорости потока $W_{\text{кол}\ r}$ тоже должны быть равны им, то есть

$$W_{\text{окр}} = W_{\text{кол}\,tg} = W_{\text{кол}\,r}. \tag{14}$$

Радиус окружности, полученный по уравнению (13), станет радиусом окружности равных скоростей, то есть

$$r = r_{\text{р.с}}, \tag{15}$$

как это показано на рис.1.

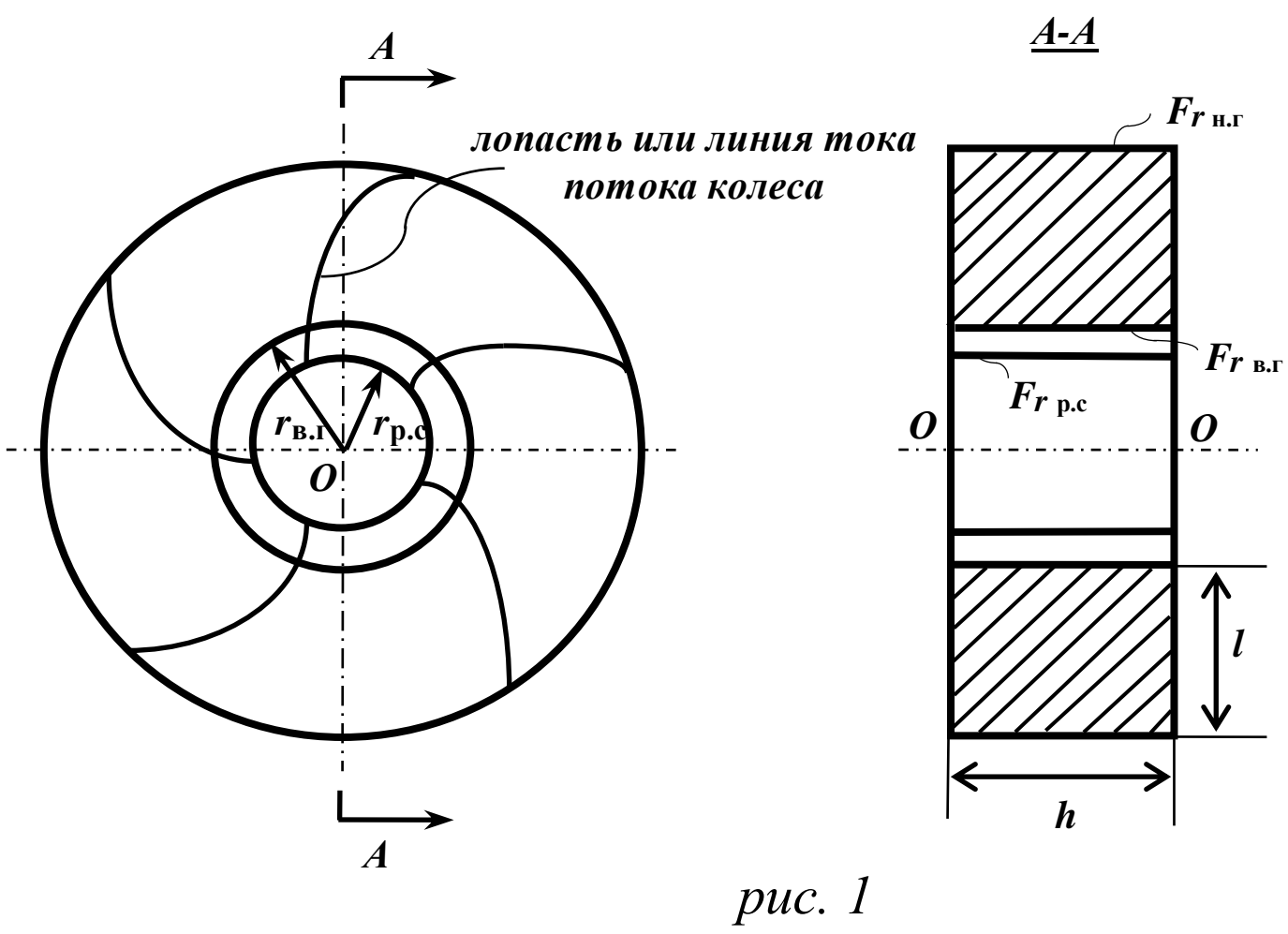

*рис. 1*

Из уравнения (14) найдем истинную величину ширины тангенциального потока $h$, или истинную ширину тангенциальной площади сечения потока $F_{tg}$ (рис. 1). Ибо при равенстве радиальных $W_{\text{кол}\ r}$ и тангенциальных $W_{\text{кол}\ tg}$ скоростей потока их площади будут равны, то есть

$$F_{tg} = F_r = F_{r_{\text{р.с}}}. \tag{16}$$

Тангенциальную площадь потока $F_{tg}$ мы нашли по уравнению (9), поэтому её величина нам известна. Радиальная площадь сечения потока $F_r$ представляет собой цилиндрическую поверхность, величина которой будет равна произведению длины окружности $2\pi r_{\text{р.с}}$, умноженной на высоту $h$ , то есть

$$F_{r_{\text{р.с}}} = 2\pi r_{\text{р.с}} h. \tag{17}$$

В уравнении (17) нам известны все величины, кроме ширины потока $h$. Радиус мы нашли по уравнению (13), а площадь сечения – по уравнению (16). Поэтому распишем уравнение (17) относительно ширины потока $h$, получим:

$$h = \frac{F_{tg}}{2\pi r_{\text{р.с}}}. \tag{18}$$

По уравнению (18) мы получили истинную величину $h$. Тогда истинной длиной тангенциальной площади сечения потока $F_{tg}$ будет длина $l$, полученная путем деления тангенциальной площади сечения потока $F_{tg}$ на ширину потока $h$, полученную по уравнению (18), то есть

$$l = \frac{F_{tg}}{h}. \tag{19}$$

Нам осталось установить размеры, а вернее, величины радиусов внутренней и наружной границ потока. Для чего нам снова придётся вернуться к балансному распределению потенциальной и кинетической энергии в потоке жидкости. Для тангенциального потока жидкости полная энергия потока $E_{tg}$ записывается в таком виде:

$$E_{tg} = V_{\text{пот}} P_{\text{ст}} + \frac{1}{2} V_{\text{пот}} \rho W^2_{\text{кол}\,tg}. \tag{20}$$

Для удобства наших пояснений отнесём полную энергию потока к единице объёма, то есть левую и правую части уравнения (20) разделим на объём потока $V_{\text{пот}}$, получим:

$$E_{\text{ед}\,tg} = P_{\text{ст}} + \frac{1}{2} \rho W^2_{\text{кол}\,tg}. \tag{21}$$

Мы знаем, что кинетическая энергия потока не должна превышать потенциальную энергию потока, то есть

$$P_{\text{ст}} \geq \frac{1}{2} \rho W^2_{\text{кол}\,tg}. \tag{22}$$

Для идеальной жидкости преобразование движения потока колеса в полную энергию потока идёт при граничных условиях распределения кинетической и потенциальной энергий потока, когда потенциальная энергия потока равна его кинетической энергии, то есть

$$P_{\text{ст}} = \frac{1}{2} \rho W^2_{\text{кол}\,tg}. \tag{23}$$

Это значит, что для тангенциального потока неравенство (22) переходит в равенство (23).

Полная энергия тангенциального потока переходит в полную энергию радиального потока, как нам известно. Скорость движения радиального потока $W_{\text{кол}.r}$ изменяется в зависимости от величины площади сечения радиального потока. Поэтому максимальная радиальная скорость тоже определяется неравенством такого вида:

$$P_{\text{ст}} \geq \frac{1}{2} \rho W^2_{\text{кол}\,r} \tag{24}$$

Так как полная энергия тангенциального потока полностью без потерь переходит в полную энергию радиального потока, то ограничительное балансное распределение потенциальной и кинетической энергии для идеальной жидкости будет одинаковым, как для радиального, так и для тангенциального потока. Это значит, что максимальные радиальные скорости потока не могут быть больше максимальных тангенциальных скоростей. По этой причине предельное ограничение максимальных радиальных скоростей, которые мы записали неравенством (24), в этом своём предельном случае может определяться равенством (23). Это равенство сохраняется в том случае, когда радиальные и тангенциальные площади сечения потока равны между собой.

Это условие выполняется на окружности равных скоростей $r_{\text{р.с}}$, или на радиальной площади сечения, образованной радиусом окружности равных скоростей. Отсюда следует, что мы не можем взять внутреннюю границу потока с радиусом меньшим, чем радиус окружности равных скоростей. Следовательно, мы можем взять радиус внутренней границы потока либо равным радиусу окружности равных скоростей $r_{\text{р.с}}$, либо б*о*льшим, чем этот радиус (рис. 1):

$$r_{\text{в.г}} \geq r_{\text{р.с}}. \tag{25}$$

Этого условия ещё недостаточно для выбора внутренней границы потока. Поэтому мы рассмотрим ещё одно условие, связанное с энергетическим распределением во входном потоке насоса.

Полная энергия потока на входе в насос нам задаётся по условию проектирования. При этом она много меньше, чем энергия потока колеса насоса. Полная энергия входного потока скоростей $E_{\text{вх.ед}}$ записывается таким уравнением:

$$E_{\text{вх.ед}} = P_{\text{ст.вх}} + \frac{1}{2}\rho W_{\text{вх}}^2. \qquad (26)$$

Для этого потока тоже существует балансное распределение потенциальной и кинетической энергий, которое определяется неравенством:

$$P_{\text{ст.вх}} \geq \frac{1}{2}\rho W_{\text{вх}}^2. \qquad (27)$$

Этим неравенством определяются максимально возможные скорости входного потока на внутренней граничной площади потока колеса насоса. Реальные скорости на этой граничной площади сечения потока $F_{r_{\text{в.г}}}$ определяются уравнением движения такого вида:

$$M_{\text{пот}} = \rho F_{r_{\text{в.г}}} W_{\text{вх}}. \qquad (28)$$

В уравнении (28) расход массы жидкости задаётся нам по условию проектирования, плотность жидкости $\rho$ тоже задаётся. Радиус внутренней граничной радиальной площади сечения потока мы определили неравенством (25). Это значит, что радиус $r_{\text{в.г}}$ этой площади сечения потока мы можем изменять в пределах неравенства (25).

Исходя из двух условий энергетического распределения в потоке колеса насоса и во входном потоке жидкости, мы принимаем необходимый радиус для внутренней границы потока $r_{\text{в.г}}$. Тем самым мы принимаем размеры внутренней границы потока, которые будут равны $r_{\text{в.г}}$ (рис. 1).

Далее определяем радиус наружной граничной поверхности потока $r_{\text{н.г}}$. Он будет равен сумме радиуса внутренней граничной поверхности $r_{\text{в.г}}$ и длине тангенциальной площади сечения потока $l$, которую мы определили уравнением (19), то есть

$$r_{\text{н.г}} = r_{\text{в.г}} + l. \qquad (29)$$

Зависимость (29) определит нам радиус, или размер, наружной граничной поверхности потока колеса.

Теперь мы получили все необходимые размеры для объёма потока колеса насоса, которые являются искомыми размерами непосредственно самого колеса насоса. Ширину колеса $h$ мы нашли по зависимости (18). Радиус внутренней граничной поверхности $r_{\text{в.г}}$ – из условия распределения энергий, а радиус наружной граничной поверхности $r_{\text{н.г}}$ – по зависимости (29).

Отметим, что любой двигатель, в том числе электрический, имеет ограничение по крутящему моменту, который определяется как диаметр, умноженный на силу, направление которой совпадает с направлением тангенциального движения. Для нашего потока таким диаметром будет средний диаметр потока между внутренней и наружной поверхностью потока. Тангенциальные динамические силы давления мы определим по зависимости:

$$F_{tg} P_{\text{дин}\,tg} = F_{tg}\,\frac{1}{2}\rho W_{\text{кол}\,tg}^2. \qquad (30)$$

Затем, умножив эту силу на средний диаметр потока, мы получим крутящий момент. Величину этого крутящего момента мы должны будем сравнить с допускаемой максимальной величиной крутящего момента выбранного нами электродвигателя. На этом заканчивается окончательный выбор электродвигателя для нашего насоса.

### *3. Нахождение размеров профиля лопастей колеса насоса*

Профиль лопасти колеса насоса должен совпадать с формой линии тока плоского установившегося потока колеса насоса. Линия тока плоского установившегося потока жидкости определяется логарифмической спиралью, зависимость которой мы записали выше как

$$r = ae^{\varphi}\,.$$

В этом уравнении $a$ есть величина постоянная, которая принимается из начальных условий построения кривой. Величина $e$ имеет постоянное значение равное 2,72. Поэтому в уравнении логарифмической спирали остаются всегда два неизвестных: это радиус $r$ и угол поворота $\varphi$. Задавая для одного из них числовое значение, мы можем получить числовое значение для другого.

Начальными условиями для построения логарифмической спирали будет точка $A$ на окружности равных скоростей, радиус которой мы определили уравнением (13) (рис. 1). Тогда значение $a$ в уравнении логарифмической спирали будет равно величине радиуса окружности равных скоростей $r_{\text{р.с}}$, то есть

$$a = r_{\text{р.с}} \text{ при } \varphi = 0,$$

так как тогда угол $\varphi$ примет начальное значение равное нулю.

Подставим эти значения в уравнение логарифмической спирали, получим:

$$r = r_{\text{р.с}} \cdot e^{0} = r_{\text{р.с}}\,,$$

поскольку $e$ в степени ноль равно единице.

Таким образом, мы определили начальное положение точки $A$ на окружности радиуса равных скоростей. Далее совмещаем начало полярной системы координат с осью потока $O$ и приступаем к построению профиля линии тока по зависимости:

$$r = r_{\text{р.с}} \cdot e^{\varphi}\,. \tag{32}$$

Затем, задаваясь углом $\varphi$ через равные небольшие промежутки, мы будем получать точки логарифмической спирали. Чем точнее нам нужно получить профиль линии тока, тем меньше мы должны брать равные промежутки угла $\varphi$. Отметим, что в этой зависимости значения угла $\varphi$ берутся не в градусах, а в радианной мере. Продолжают построение точек кривой до внешней границы потока. Затем эти точки соединяют плавной кривой. Отрезок этой кривой, который находится в пределах внутренней и наружной границ потока, даст нам профиль искомой лопатки колеса насоса. Объём лопастей колеса должен учитываться в объёме потока насоса. На этом заканчивается поиск необходимых размеров для проточной полости колеса проектируемого насоса. <…>

***

от редактора: **Отрывок из книги Г. Галилея «Пробирных дел мастер»:**

«К сказанному надлежит добавить ещё одно измышление, против которого Сарси, по его словам, не имеет возражений, и даже называет его тождеством. По его мнению, утверждать, что жидкое тело, заключённое в полости сферического твёрдого тела, может быть увлечено последним, если твёрдое тело приведено во вращение, означает утверждать столь же легко и естественно, что и твёрдое тело, заключённое в жидкую субстанцию, было бы вовлечено в движение, если бы жидкость вращалась.

Утверждать такое – всё равно, что думать, будто подобно тому, как судно переносится и увлекается течением реки, так и вода в стоячем пруду должна увлекаться вслед за движением лодки, что совершенно неверно. Прежде всего, сошлёмся на опыт: мы знаем, что течение реки приводит в движение судно или даже тысячу судов, заполняющих всю реку, в то время как судно, с какой бы скоростью оно ни плыло, не увлекает за собой воду. <…> Так тяжёлый груз, подвешенный на веревке, сохраняет в течение многих часов импет и движение, переданные ему одним толчком; вместе с тем можно сколь угодно сильно привести в движение воздух в комнате, но, как только импет *(impetus (лат.) – побудительная сила движения, пояснение взято из книги – ред.*), вызвавший движение, перестает действовать воздух полностью успокаивается, и переданное ему движение бесследно исчезает.

Если окружающее и движущееся тело – жидкость, и оно действует с силой на находящееся в нём твёрдое, массивное и тяжёлое тело, то такое тело передаёт своё движение объекту, который в силу своей природы особенно пригоден для получения и сохранения движения в течение долгого времени. Это означает, что второй импет, идущий вслед за первым, налагается на движение, уже переданное твёрдому телу; третий импет налагается на движение, переданное телу первым и вторым импетом; четвертый импет добавляется к действию, произведенному первым, вторым и третьим импетом; так, шаг за шагом,  движение движимого тела не только сохраняется, но и получает приращение. Но если движимое тело – жидкость, легкая и тонкая и, следовательно, не сохраняющая, а тот час же утрачивающая передаваемое ей движение, то пытаться сообщить ей скорость – занятие столь же бесплодное, как пытаться наполнить бочку Данаид, которая мгновенно опорожняется, стоит лишь её наполнить водой. Смотри же, синьор Лотарио (*имя Сарси. – ред.*), сколь велика разница между двумя случаями, которые ты считал тождественными».

(Г. Галилей «Пробирных дел мастер» (1623 г.))

***

По всем вопросам можно обращаться к редактору по адресу:

arodionova1@rambler.ru